\documentclass[12pt,letterpaper]{article}
\usepackage[utf8]{inputenc}
\usepackage[english]{babel}
\usepackage[vmargin=0.6in,hmargin={1in,1in},headsep=0.3in,headheight=1in]{geometry}
\usepackage{subfiles}
\usepackage{tikz}
\usepackage{fix-cm}
\usepackage{enumitem}
\usepackage{multicol}
\usepackage{amsmath}
\usepackage{amsthm,amssymb,amsfonts}
\usepackage{array}
\usepackage{relsize}
\usepackage{xcolor}
\usepackage{lastpage}
\usepackage{blindtext}
\usepackage[linktocpage=true]{hyperref}
\hypersetup{
	colorlinks,
	citecolor=blue,
	filecolor=black,
	linkcolor=blue,
	urlcolor=blue
}
\usepackage{graphicx}
\usepackage{subcaption}
\usepackage{wrapfig}
\usepackage{indentfirst}
\usepackage{hyperref}
\usepackage{soul}
\usepackage{empheq}
\usepackage{placeins}
\usepackage{mathabx}
\setlist{parsep=0pt,listparindent=\parindent}
\setlist[itemize]{noitemsep, topsep=0pt}
\setlist[enumerate]{noitemsep, topsep=0pt}
\setlist{parsep=0pt,listparindent=\parindent}

\stepcounter{footnote}
\graphicspath{{Images/}{../Images/}}
\catcode`\^=13\def^#1{\sp{#1}{}}
\catcode`\_=13\def_#1{\sb{#1}{}}

\title{\textbf{\Huge{Curvature Invariants for the Accelerating Nat\'ario Warp Drive}}}
\author{B.~Mattingly$^{1,2}$\footnote{mailto: \em\texttt{\href{Brandon_Mattingly@Baylor.edu}{Brandon\_Mattingly@Baylor.edu}}}, A.~Kar$^{1,2}$, M.~Gorban$^{1,2}$, W.~Julius$^{1,2}$, C.~K.~Watson$^{1,2}$, \\
M.~D.~Ali$^{1,2}$, A.~Baas$^{1,2}$, C.~Elmore$^{1,2}$, J.~S.~Lee$^{1,2}$, B. Shakerin$^{1,2}$, \\
E.~W.~Davis$^{1,3}$ and G.~B.~Cleaver$^{1,2}$}
\date{}
\begin{document}
		
\maketitle
\vspace{-0.8cm}
\begin{center}
\begin{minipage}[c]{0.72\textwidth}
	$^1$\emph{Early Universe, Cosmology and Strings \textnormal{(EUCOS)} Group, Center for Astrophysics, Space Physics and Engineering Research \textnormal{(CASPER)}, Baylor University, Waco, TX 76798, USA}\\ \\
	$^2$\emph{Department of Physics, Baylor University, Waco, TX 76798, USA}\\ \\
	$^3$\emph{Institute for Advanced Studies at Austin, 11855 Research Blvd., Austin, TX 78759, USA}\\
\end{minipage}
\end{center}	

\begin{minipage}{0.92\textwidth}
	\textbf{\large{Abstract}:} A process for using curvature invariants is applied to evaluate the accelerating Nat\'ario warp drive. 
	Curvature invariants are independent of coordinate bases and plotting the invariants is free of coordinate mapping distortions.
	While previous works focus mainly on the mathematical description of the warp bubble, plotting curvature invariants provides a novel pathway to investigate the Nat\'ario spacetime and its characteristics. 
    For warp drive spacetimes, there are four independent curvature invariants the Ricci scalar, $r_1$, $r_2$, and $w_2$.
	The invariant plots demonstrate how each curvature invariant evolves over the parameters of time, acceleration, skin depth and radius of the warp bubble. 
	They show that the Ricci scalar has the greatest impact of the invariants on the surrounding spacetime.
	They also reveal key features of the Nat\'ario warp bubble such as a flat harbor in the center of it, a~dynamic wake, and the internal structures of the warp bubble.
	\end{minipage}
	
\section{Introduction}
    In Newtonian mechanics and special relativity, the~velocity of any particle is fundamentally constrained by the speed of light, $c$.
    No particle can move through spacetime at a velocity greater than $c$ in any reference frame.
    However, general relativity allows a global superluminal velocity while retaining a subluminal local velocity.  
    Alcubierre demonstrated a solution for the Einstein field equations that allowed a spaceship to make a trip to a distant star in an arbitrarily short proper time~\cite{Alcubierre:1994tu}. 
    He proposed a warped spacetime that pairs a local contraction of spacetime in front of the spaceship with a local expansion of spacetime behind the ship. 
    While the spaceship remains within its own light cone and never exceeds $c$, globally the relative velocity, which is defined as proper spatial distance divided by proper time, may be much greater than $c$. 
    \mbox{Based on this principle,} he named the faster-than-light (FTL) propulsion mechanism a ``warp drive.''
    
    FTL travel requires eight general conditions for a spaceship to carry a passenger safely~\cite{Davis}. First,~the~rocket equation is not required for travel by way of the warp. 
    Second, the~travel time through the FTL space warp to a distant star may be reduced to less than one year as seen both by the passengers in the warp and by stationary observers outside the warp. 
    Third, the~proper time as measured by the passengers should not be dilated by any relativistic effects. 
    Fourth, any~tidal-gravity accelerations acting on passengers inside of the spaceship may be reduced to be less than the acceleration of gravity near the Earth's surface, $g_{\Earth}$. 
    Fifth, the~local speed of the spaceship must be less than $c$.
    Sixth, the~matter of the spaceship must not couple with any exotic material needed to generate the FTL space warp. 
    Seventh, the~FTL space warp should not generate an event horizon.
    \mbox{Finally, the~passengers} should not encounter a singularity inside or outside of the FTL warp.

    Traversable wormholes and warp drives are two known example spacetimes that satisfy these eight requirements~\cite{Natario:2001tk,Alcubierre:1994tu,Krasnikov:1995ad,VanDenBroeck:1999sn,5,6,9,11,Loup,Loup2}. 
    While they are mathematical solutions to Einstein's equations, building such devices are not achievable in the foreseeable future due to severe engineering constraints. 
    \mbox{In~\cite{3}, the~authors} used the method of calculating and plotting curvature invariants to analyze several types of wormholes. 
    In the present work, the~authors will adapt this methodology to analyze the accelerating Nat\'ario Warp Drive.

    Since Alcubierre, there has been considerable research into FTL warp drives. Krasnikov~considered a non-tachyonic FTL warp bubble and showed it to be possible mathematically~\cite{Krasnikov:1995ad}. 
    Van~Den Broeck modified Alcubierre's warp drive to have a microscopic surface area and a macroscopic volume inside. 
    He showed that the modification allowed a warp bubble with more reasonable negative energy requirements of a few solar masses and that it has a more lenient violation of the null-energy-conditions (NEC).
    Later, Nat\'ario improved upon Alcubierre's work by presenting a warp drive metric such that zero spacetime expansion occurs~\cite{Natario:2001tk}. 
    Instead of riding a contraction and expansion of spacetime, the~warp drive may be observed to be ``sliding'' through the exterior spacetime at a constant global velocity. 
    Loup expanded Nat\'ario's work to encompass a global acceleration~\cite{Loup,Loup2}.
    Finally, recent research has investigated the Einstein equations for the Alcubierre spacetime with a dust matter distribution as its source~\cite{Santos-Pereira:2020puq}. 

    While the mathematics of a warp drive is well developed, mapping the spacetime around the warp drive remains unexplored until recently.
    Considering that a ship inside of a warp bubble is causally disconnected from the exterior, computer simulations of the surrounding spacetime are critical for the ship to map its journey and steer the warp bubble~\cite{Natario:2001tk}.
    Alcubierre used the York time, which is defined as $\Theta=\frac{v_s}{c}\frac{x-x_s}{r_s}\frac{df}{dr_s}$, to map the volume expansion of a warp drive~\cite{Alcubierre:1994tu}. 
    He plotted the York time to show how spacetime warped behind and in front of the spaceship. 
    While the York time is appropriate when the 3-geometry of the hypersurfaces is flat, it will not contain all information about the curvature of spacetime in non-flat 3-geometries such as the accelerating Nat\'atio warp drive spacetime. 
    Alternatively, plotting the curvature invariants for a warp drive will display the spacetime's curvature. 

    Christoffel proved that scalars constructed from the metric and its derivatives must be functions of the metric itself, the~Riemann tensor, and its covariant derivatives~\cite{Chris}. 
    In particular, curvature invariants are scalar products of Riemann, Ricci or Weyl tensors, or their covariant derivatives. 
    Fourteen curvature invariants in $(3+1)$ dimensions have been defined in the literature, but the total rises to seventeen when certain non-degenerate cases are taken into account~\cite{ZM}. 
    Carminati and McLenaghan (CM) demonstrated a set of invariants that had several attractive properties.
    Their invariant set maintains general independence, requires each invariant to be of lowest possible degree, and contains a minimal independent set for any Petrov type and choice of Ricci tensor~\cite{CM}.
    An invariant is considered to be independent if it cannot be written in terms of other members of the set that are of equal or lower degree. 
    A set of invariants is independent if each element of the set is an independent invariant. The polynomial relationships, called syzgies, between the individual elements in a set of invariants may be used to determine independence. 
    The syzgies of a set of invariants may be determined algorithmically.
    For the case of Class B spacetimes, the syzgies between the invariants further reduces the independent set of invariants to be only the four specific ones: $R$, $r_1$, $r_2$ and $w_2$~\cite{Santosuosso:1998he}. 
    Curvature invariants are the same regardless of your choice of coordinate.
    Consequently, a~complicated spacetime, such as that of a warp drive, may be displayed without distortion when its curvature invariants are plotted.

    Recent research by Overduin et al.~\cite{Henry} computed and plotted a set of independent curvature invariants for the hidden interiors of Kerr-Newman black holes. 
    It produced visually stunning 3D plots of the surprisingly complex nature of spacetime curvature in Kerr-Newman black hole interiors.
    In addition, the~calculation of curvature invariants in black holes is a rich topic of recent study~\cite{14,Baker,Abdelqader:2014vaa,13,16}.
    This work motivated the present authors to undertake a similar study for the Nat\'ario warp drive.
    While the individual invariant functions are colossal in size and require a lengthy time to calculate, their plots can be quickly scanned and understood.

\section{Method to Compute the Invariants}
	The CM curvature invariants can be calculated from any given line element.
    From the line element, the metric $g_{ij}$ is identified with the indices $\{i,j,k,l\}$ ranging from $\{0,n-1\}$, where $n$ is the number of spacetime dimensions.
    The accelerating Nat\'ario warp drive was derived using $n=4$ for its dimensions\cite{Loup}.
    In spacetime, a complex null tetrad may be identified from a given line element.
	In this paper, a null tetrad $(l_i,k_i,m_i,\bar{m}_i)$ is found for the accelerating Nat\'ario metric. 
	It is emphasized that the scalar invariants are independent of both the choice of coordinate parameterization and the choice of the null tetrad.
	A different choice of coordinates for each spacetime's tetrad will result in invariants related by algebriac expressions to the ones plotted in this paper.
	From the metric, the affine connection $\Gamma^i_{jk}$, Riemann tensor $R^i_{jkl}$, Ricci tensor $R_{ij}$, Ricci scalar $R$, trace free Ricci tensor $S_{ij}$ and Weyl tensor $C_{ijkl}$ are calculated. 
    The Newman-Penrose (NP) curvature components may be computed from the null tetrad, the trace-free Ricci tensor, and the Weyl tensor \cite{Stephani:2003tm}. 
	The thirteen different CM invariants are defined in \cite{CM}. 
	Only four of these invariants are required by the syzgies for Class B spacetimes: the Ricci scalar \eqref{eq:R}, the first two Ricci invariants \eqref{eq:r1} \eqref{eq:r2}, and the real component of the Weyl invariant J \eqref{eq:w2} \cite{Santosuosso:1998he}. 
	In terms of the NP curvature coordinates, the CM invariants are:
	\begin{align}
	R &= g_{ab} R^{ab}, \label{eq:R} \\
	\begin{split}
	r_1& = \frac{1}{4} S_a^b S_b^a \\ & = 2\Phi_{20}\Phi_{02}+2\Phi_{22}\Phi_{00}-4\Phi_{12}\Phi_{10}-4\Phi_{21}\Phi_{01}+4\Phi_{11}^2, \end{split} \label{eq:r1}
	\\
	\begin{split}
	r_2& = -\frac{1}{8} S_a^b S_c^a S_b^c \\ & = 6\Phi_{02}\Phi_{21}\Phi_{10}-6\Phi_{11}\Phi_{02}\Phi_{20}+6\Phi_{01}\Phi_{12}\Phi_{20}-6\Phi_{12}\Phi_{00}\Phi_{21} -6\Phi_{22}\Phi_{01}\Phi_{10}+6\Phi_{22}\Phi_{11}\Phi_{00}, \end{split} \label{eq:r2}
	\\
	w_2& = -\frac{1}{8} \bar{C}_{a b c d} \bar{C}^{a b e f} \bar{C}^{c d}{}_{e f} \nonumber \\ & = 6\Psi_4\Psi_0\Psi_2-6\Psi_2^3-6\Psi_1^2\Psi_4-6\Psi_3^2\Psi_0+12\Psi_2\Psi_1\Psi_3. \label{eq:w2}
	\end{align}
	The tetrad components of the traceless Ricci Tensor are $\Phi_{00}$ through $\Phi_{22}$~\cite{Stephani:2003tm}.
	The complex tetrad components $\Psi_0$ to $\Psi_5$ are the six complex coefficients of the Weyl Tensor due to its tracelessness.

\section{Warp Drive Spacetimes}
    Alcubierre and Nat\'ario developed warp drive theory using $(3+1)$ dimensional ADM formalism. 
    Spacetime is decomposed into space-like hypersurfaces parametrized by the value of an arbitrary time coordinate $dx^0$ \cite{ADM,Marqu}.
    Two nearby hypersurfaces, $x^0=$\ constant and $x^0+dx^0=$\ constant, are separated by a proper time $d\tau=N(x^\alpha, x^0)dx^0$. 
    The ADM four-metric is 
    \begin{equation}
	g_{ij}=
	\begin{pmatrix}
	\ -N^2-N_\alpha N_\beta g^{\alpha\beta}&\ \ N_\beta \ \\
	\ N_\alpha&\ \ g_{\alpha\beta} \ \
	\end{pmatrix}, \label{eq:ADM}
	\end{equation}
	where $N$ is the lapse function, $N_\alpha$ is the shift vector between hypersurfaces, and $g_{\alpha\beta}$ is the $3$-metric of the hypersurfaces. 
	A warp drive  is defined as a globally hyperbolic spacetime $(M, g)$, where $M=\mathbb{R}^4$ and $g$ is given by the line element
    \begin{equation}
        \text{d}s^2=-\text{d}t^2+\sum_{i=1}^{3}(\text{d}x^i-X^i \text{d}t)^2, \label{eq:6}
    \end{equation}
    for three unspecified bounded smooth functions $(X^i)=(X,Y,Z)$ in Cartesian coordinates \cite{Natario:2001tk,Alcubierre:1994tu}. 
    The functions form a vector field given by $\textbf{X}=X^i\frac{\partial}{\partial x^i}=X\frac{\partial}{\partial x}+Y\frac{\partial}{\partial y}+Z\frac{\partial}{\partial z}$. $\textbf{X}$ is a time-dependent vector field in Euclidean 3-space. 
    Class B$1$ spacetimes include all spherical, planar and hyperbolic spacetimes \cite{Santosuosso:1998he}.
    Since the general warp drive spacetime is globally hyperbolic, any specific choice of $X^i$ will be a class B$1$ spacetime, and the complete set of  curvature invariants will be Eqs.~\eqref{eq:R} through \eqref{eq:w2}.
    
    Nat\'ario considered a specific choice of $N$ and $N_\alpha$ such that a net expansion/contraction of the surrounding spacetime is not necessary at all \cite{Natario:2001tk}.
    Instead, a volume-preserving warp drive "slides" the warp bubble region through space where the space in front of it is expanded and balanced by an opposite contraction of the space behind it.
    Originally, Nat\'ario only considered the warp bubble to slide at a constant superluminal velocity.
    Later, six line elements for the Nat\'ario spacetime metric with a constant acceleration were derived \cite{Loup}. 
    The specific equation for the Nat\'ario warp drive line element in the parallel covariant $(3+1)$ dimensional ADM is
    \begin{equation}
        \text{d}s^2=\big(1-2 X_t+(X_t)^2-(X_{r_s})^2-(X_{\theta})^2\big)\text{d}t^2+2\big(X_{r_s}\text{d}r_s+X_{\theta} r_s \text{d}\theta\big)\text{d}t-\text{d}r_s^2-r_s^2\text{d}\theta^2-r_s^2\sin^2{\theta}\text{d}\phi^2. \label{eq:metric}
    \end{equation}
    which is in the spherical coordinates, for which $0\le r_s <\infty;\ 0\leq\theta\leq\pi;\ 0\leq\varphi\leq 2\pi$, $-\infty< t<\infty$,  and $a$ is the constant acceleration.
    The covariant shift vector components are given by 
    \begin{align}
        X_t&=2n(r_s) a r_s \cos{\theta}, \label{eq:Xt} \\
        X_{r_s}&=2[2n(r_s)^2+r_sn'(r_s)]at\cos{\theta}, \label{eq:Xrs} \\
        X_{\theta}&=-2n(r_s)at[2n(r_s)+r_sn'(r_s)]r_s^2\sin{\theta}. \label{eq:Xth}
    \end{align} \\
    The Nat\'ario warp drive continuous shape function is 
    \begin{equation}
        n(r_s)=\frac{1}{2}\Bigg[1-\frac{1}{2}\Big(1-\tanh[\sigma(r_s-\rho)]\Big)\Bigg]. \label{eq:nrs}
    \end{equation}
    where $\sigma$ is the skin depth of the warp bubble, and $\rho$ is the radius of the warp bubble. 
    Appendix \ref{app:null} derives the comoving null tetrad for Eq.~\eqref{eq:metric}. It is 
    \begin{align} \label{eq:null}
        l_i&=\frac{1}{\sqrt{2}}\begin{pmatrix}1-X_t+X_{r_s}\\-1\\0\\0\end{pmatrix}, &
        k_i&=\frac{1}{\sqrt{2}}\begin{pmatrix}1-X_t-X_{r_s}\\1\\0\\0\end{pmatrix}, \nonumber\\ \\
        m_i&=\frac{1}{\sqrt{2}}\begin{pmatrix}X_\theta\\0\\-r\\i r \sin{\theta}\end{pmatrix}, &  
        \bar{m}_i&=\frac{1}{\sqrt{2}}\begin{pmatrix}X_\theta\\0\\-r\\-i r \sin{\theta}\end{pmatrix}. \nonumber
    \end{align}
    The comoving null tetrad describes light rays traveling parallel to the warp bubble.
    The four CM invariants in Eqs.~\eqref{eq:R} through \eqref{eq:w2} may be derived from Eqs.~\eqref{eq:metric} and \eqref{eq:null}.

\section{Invariants for the Accelerating Nat\'ario Warp Drive}
    Figure~\ref{fig:1} below shows how the Ricci scalar evolves from $t=0$ s to $t=100$ s, while setting the other free parameters to $\rho=100$ m, $\sigma=50000 \ \text{m}^{-1}$, and $a=1.0 \ \text{m s}^{-2}$.
    In addition, the~plots of the invariants $r_1$, $r_2$, and $w_2$ are included in Appendix \ref{app:fig} in Figure~\ref{fig:2} through Figure~\ref{fig:4}. 
    The~figures provide rich details of the features in and around the warp bubble. 
    Each plot has a safe harbor within $\rho\leq100$ m where the invariant's magnitude is zero and the curvature is flat. 
    This observation is consistent with a spaceship traveling at a velocity less than $c$ in the interior of the harbor and experiencing only flat space throughout the entire time evolution.
    A symmetrical wake lies on both sides of the harbor and is characterized by large positive or negative magnitudes of the invariant. 
    The~warp bubble travels perpendicular to the wake from left to right in the plots.
    Its motion is driven by the constant curvature outside the harbor in the Ricci scalar.
    Initially, the~curvature invariant has a small positive value in front of the harbor and a small negative value behind it as can be seen in Figure~\ref{fig:1}a.
    Afterwards, a~constant negative curvature lies in front and behind of the harbor and wake.
    While the form of the Ricci scalar's plots quickly reaches a constant shape, the~Ricci scalar's magnitude increases  \mbox{as time advances.}
    
\subsection{Invariant Plots While Varying the Time} \label{ss:Time}

    Choosing the same values for $\rho$, $\sigma$, and $a$ as the Ricci scalar and varying the time, Figure~\ref{fig:2} in Appendix \ref{app:fig} shows the time evolution for the $r_1$ invariant. 
    It has many similar features to the Ricci scalar. 
    It contains the safe harbor, a~wake running perpendicular to the direction of motion, and its magnitude increases as time advances. 
    The~first difference is its positive magnitude and lack of internal structure in the wakes. 
    The~wake increases subtly in angular size as time increases.
    Finally,~the~space has no curvature in front and behind of the harbor and wake.
    
    The~invariant $r_2$ shares the same basic properties of the Ricci scalar and $r_1$: the safe harbor, a~wake, is asymptotically flat away from the bubble, and a linear increase with time. 
    It is similar in shape to $r_1$, but it increases in magnitude more drastically and has the same internal structure as the Ricci scalar. 
    
    The~invariant $w_2$ has the same features as that of the Ricci scalar, $r_1$, and $r_2$: the safe harbor, a~wake, and an increase in magnitude as time advances.
    Outside of the harbor and wake, it is asymptotically flat  like the invariants $r_1$ and $r_2$.
    The~invariant's wake contains a small amount of internal structure at lower time values, but as time reaches $100 \ s$, the~internal structures begins to form crenulations. 
    As~time continues to evolve, the~crenulations travel out parallel to the length of the wake from the center of warp bubble. 
    It can be speculated that this would cause an erratic flight path of the bubble as the crenulations' ripple outwards. 
    It can also be speculated that the interior structure of the Ricci scalar and invariant $r_2$'s wake would exhibit similar behavior at higher time values.
    
    The~shape of each invariant agrees with Nat\'ario's description of the warp bubble sliding through spacetime.
    The~positive and negative curvature from the wake will transport a ship in the harbor. 
    The~Ricci scalar has the greatest impact on the spacetime outside of the harbor and wake due to the constant curvature in this area.
    Each invariant's magnitude experiences an increase as time advances.
    The~invariant $w_2$ has the greatest impact on the wake as its magnitude increases the most as time advances.
    The~wake's shape also shows that there is internal structure to the warp bubble.
    Theoretically, the~crenulations exhibited in the invariant $w_2$ would make navigation challenging. 
    But, they can be avoided by accelerating the warp drive for only short periods.
    The~increase suggests an engineering constraint on a maximum achievable global velocity. 
    
     \begin{figure}[ht]
    	\begin{subfigure}{.45\linewidth}
    	    \centering
    		\includegraphics[scale=0.25]{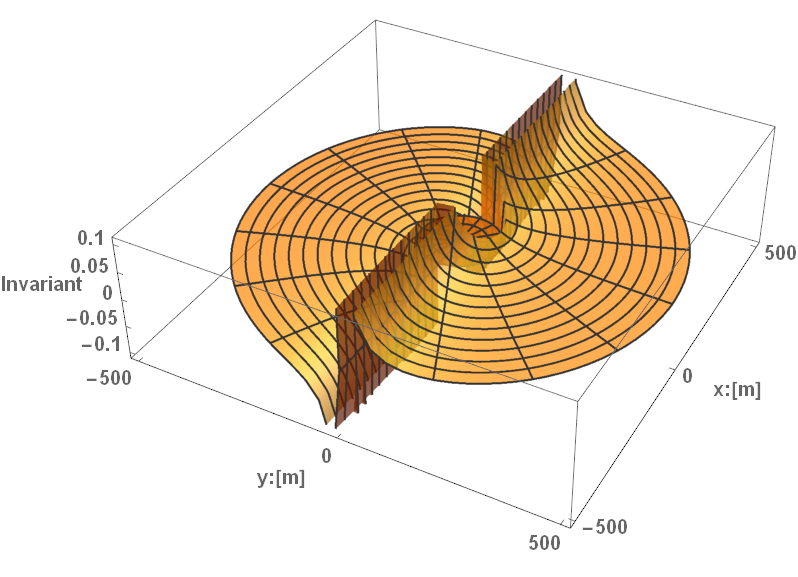}
    		\caption{$t=0.0$ s}
    		\label{Rs50000p100a1t0}
    	\end{subfigure}
    	~
    	\begin{subfigure}{.45\linewidth}
    	    \centering
    		\includegraphics[scale=0.25]{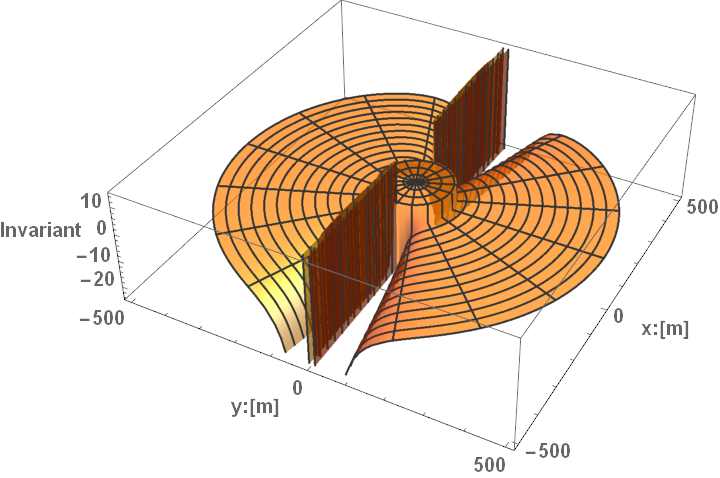}
    		\caption{$t=1.0$ s}
    		\label{Rs50000p100a1t1}
    	\end{subfigure}
    	\par\bigskip
    	\begin{subfigure}{.45\linewidth}
    	    \centering
    		\includegraphics[scale=0.25]{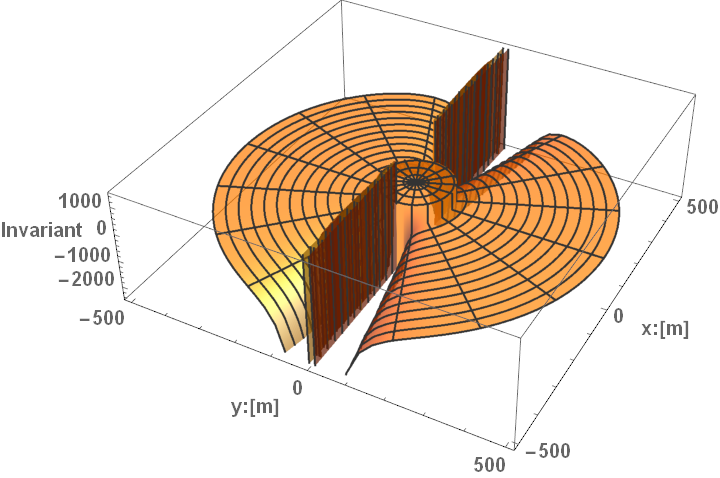}
    		\caption{$t=10.0$ s}
    		\label{Rs50000p100a1t10}
    	\end{subfigure}
    	~
    	\begin{subfigure}{.5\linewidth}
    	    \centering
    		\includegraphics[scale=0.25]{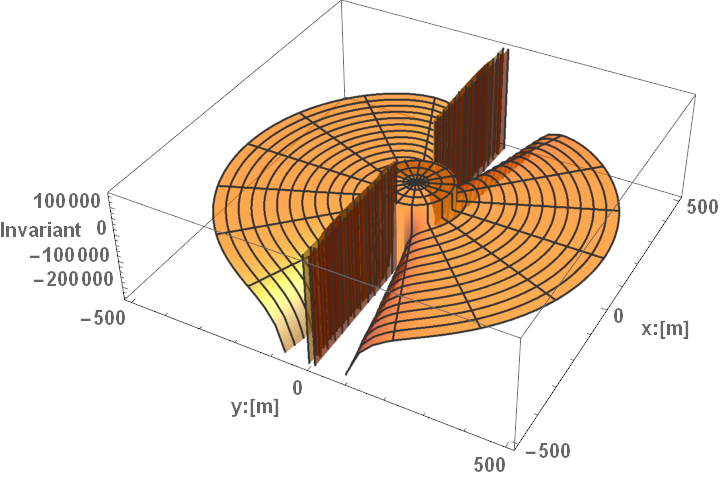}
    		\caption{$t=100.0$ s}
    		\label{Rs50000p100a1t100}
    	\end{subfigure}
    	~
    	\vspace{.25cm}
    	~
    	\caption{The time, $t$, of the Ricci Scalar $R$ is plotted in the figures above.  
    	The other parameters were set to be $\sigma=50,000 \ \text{m}^{-1}$, $\rho=100$ m, and $a=1.0 \ \text{m s}^{-2}$.
    	The plots were rendered using Mathematica's\textsuperscript{\textregistered} RevolutionPlot3D.
	    It plots a function by rotating the function around the z-axis.
	    The~amount that the space diverges from being flat is represented by the invariant function's magnitude, and its amount is labeled on the vertical axis.
	    The~x and y-axis are displayed on the plots, and they display the distance from a spaceship in the flat portion at the center of each figure.
        Each radial line in the plot corresponds to a distance of approximately $33$ m.
        The~portion of small negative curvature to the right of the harbor is to the front of the warp bubble, and the portion to the left of the harbor is the back of \mbox{the warp bubble.}} \label{fig:1}
    \end{figure}

\subsection{Invariant Plots While Varying the Acceleration}
       
    Varying the acceleration of the invariants speeds up or slows down the time evolution in the previous section. 
    Setting $\rho=100$ m, $\sigma=50000 \ \text{m}^{-1}$, and $t=1.0$ s, Figure~\ref{fig:5} below shows the variation of the acceleration of the Ricci scalar.
    The~first plot of $a=0 \ \text{m s}^{-2}$ is consistent with the lapse function being zero for any given time period. 
    The~distance between hypersurfaces will be constant. 
    The~space will be flat and no warp bubble will form as is shown. 
    The~second plot corresponds to a time slice between Figure~\ref{fig:1}a,b. 
    Similarly, the~third plot is identical to Figure~\ref{fig:1}c, and the fourth plot corresponds to a time slice after Figure~\ref{fig:1}d. 
    It can be concluded that modifying the acceleration parameter corresponds with modifying the rate of change of the hypersurfaces. 
    Additionally, our~analysis of the invariant's shape and properties in Section~\ref{ss:Time} holds for the acceleration plots.
    
    The~plots of the invariants $r_1$, $r_2$ and $w_2$ follow a similar process as the Ricci scalar. 
    They are plotted in Figure~\ref{fig:6} through Figure~\ref{fig:8} in Appendix \ref{app:fig}. When $a=0$, their plots are identical to Figure~\ref{fig:5}a. 
    Then, they can be seen as additional time slices between the plots shown in Figure~\ref{fig:2}a,b for $r_1$ and Figure~\ref{fig:3}c,d for $r_2$.  
    Some additional features are present in the plots. 
    The~invariants do warp themselves much more significantly and non-symmetricly than their counterparts for the Ricci scalar, as can be seen in Figure~\ref{fig:7}c. 
    
     \begin{figure}[ht]
	\centering
	\begin{subfigure}[t]{.47\linewidth}
	    \centering
		\includegraphics[scale=0.29]{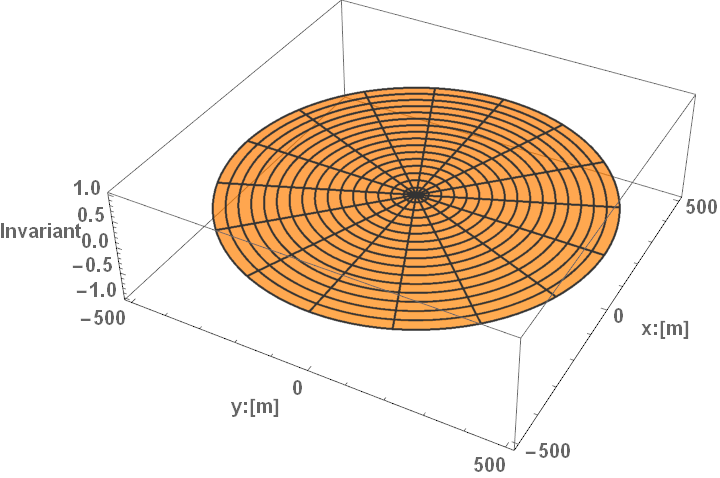}
		\caption{$a=0.0 \ \text{m s}^{-2}$}
		\label{Rs50000p100a0}
	\end{subfigure}
	~
	\begin{subfigure}[t]{.47\linewidth}
	    \centering
		\includegraphics[scale=0.29]{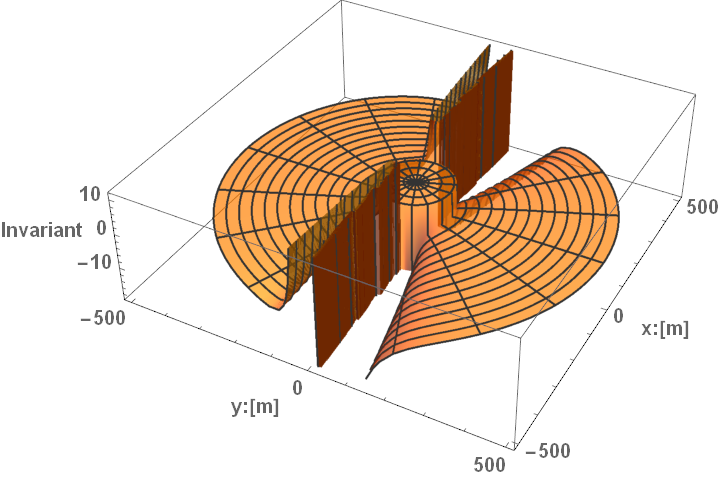}
		\caption{$a=0.1 \ \text{m s}^{-2}$}
		\label{Rs50000p100ap1t1}
	\end{subfigure}
	\par\bigskip
	\begin{subfigure}[t]{.47\linewidth}
	    \centering
		\includegraphics[scale=0.29]{Images/R/Rs50000p100a1t1.png}
		\caption{$a=1.0 \ \text{m s}^{-2}$}
		\label{Rs50000p100a1t1 2}
	\end{subfigure}
	~
	\begin{subfigure}[t]{.47\linewidth}
	    \centering
		\includegraphics[scale=0.29]{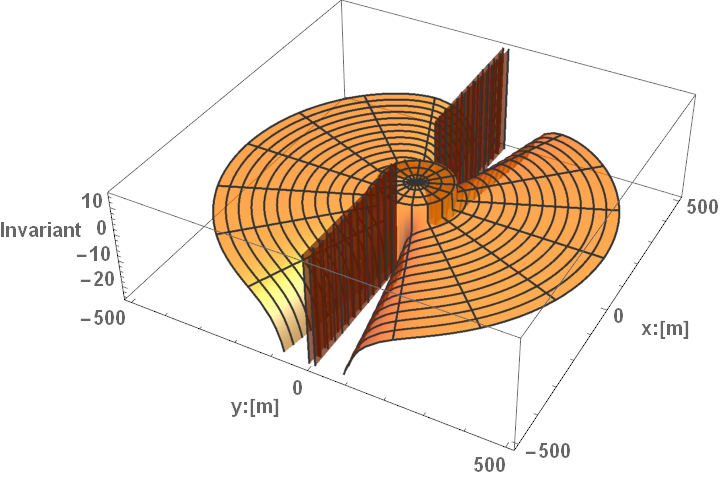}
		\caption{$t=10.0 \ \text{m s}^{-2}$}
		\label{Rs50000p100a10t1}
	\end{subfigure}
	~
    \vspace{.25cm}
    ~
	\caption{The acceleration, $a$, of the Ricci Scalar $R$ is plotted in the figures above. 
	The other parameters were set to be $\sigma=50,000 \ \text{m}^{-1}$, $\rho=100$ m, and $t=1$ s.
	The plots were rendered using Mathematica's\textsuperscript{\textregistered} RevolutionPlot3D.
	It plots a function by rotating the function around the z-axis.
	The amount that the space diverges from being flat is represented by the invariant function's magnitude, and its amount is labeled on the vertical axis.
	The x and y-axis are displayed on the plots, and they display the distance from a spaceship in the flat portion at the center of each figure.
    Each radial line in the plot corresponds to a distance of approximately $33$ m.
    The~portion of small negative curvature to the right of the harbor is to the front of the warp bubble, and the portion to the left of the harbor is the back of \mbox{the warp bubble.}
    } \label{fig:5}
    \end{figure}

\subsection{Invariant Plots While Varying the Skin Depth of the Warp Bubble}
    
    Varying the skin depth of the warp bubble, $\sigma$, does not noticeably affect the invariant plots.  
    Figure~\ref{fig:4.3} below presents the plots for each of the four invariants while doubling the skin depth and setting $\rho=100$ m, $t=1.0$ s, and $a=1.0 \ \text{m} s^{-2}$. 
    Figure~\ref{fig:9} in Appendix~\ref{app:fig} presents the remaining invariants $r_1$, $r_2$, and $w_2$, which have similar features to the Ricci Scalar.
    The~shape of the invariant does not change in the figures after doubling.
    In the Nat\'ario line element, the~skin depth occurs in $\tanh{\sigma(r_s-\rho)}$ of \eqref{eq:nrs}.
    $\sigma$ moderates the rate at which the top hat function converges to be $0$ outside the warp radius and $1$ along the depth of warp's edge near $\rho$.
    For the relatively high values of $\sigma$ plotted, the~top hat function must approach $1$, and the overall depth of the warp bubble must be very small.
    If~$\sigma$ was plotted on the order of $1 \ \text{m}^{-1}$, a~greater effect on the invariants would be observed than what is shown with the attached plots.

\begin{figure}[ht]%
	\begin{subfigure}{.45\linewidth}
	    \centering
		\includegraphics[scale=0.25]{Images/R/Rs50000p100a1t1.png}
		\caption{Ricci scalar with $\sigma=50,000 \text{m}^{-1}$}
		\label{Rs50000p100a1t1 s}
	\end{subfigure}
	~
	\begin{subfigure}{.45\linewidth}
    	\centering
		\includegraphics[scale=0.25]{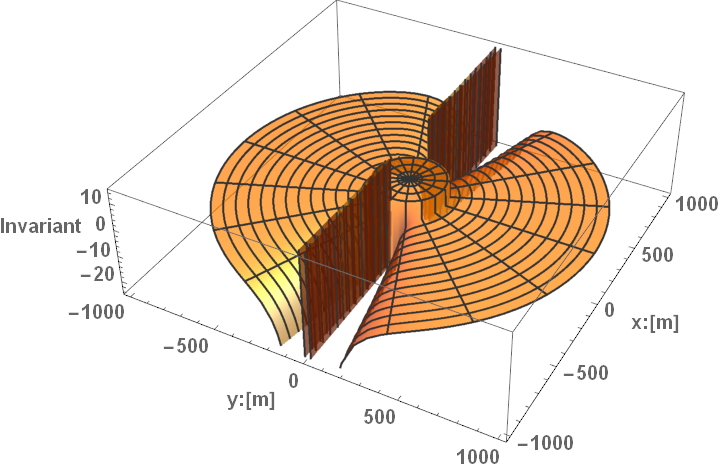}
		\caption{Ricci scalar with $\sigma=100,000 \text{m}^{-1}$}
		\label{Rs100000p100ap1t1 s}
	\end{subfigure}
	~
	\vspace{.25cm}
	~
	\caption{The warp bubble skin depth, $\sigma$, of the Ricci Scalar $R$ is plotted in the figures above.
	The~other parameters were set to be $\rho=100$ m, $a=1.0 \ \text{m s}^{-2}$, and $t=1$ s.
	The~plots were rendered using Mathematica's\textsuperscript{\textregistered} RevolutionPlot3D.
	It plots a function by rotating the function around the Z-axis.
	The~amount that the space diverges from being flat is represented by the invariant function's magnitude, and its amount is labeled on the vertical axis.
	The~X and Y-axis are displayed on the plots, and they display the distance from a spaceship in the flat portion at the center of each figure.
    Each radial line in the left hand column corresponds to a distance of approximately $33$ m.
    The~constant portion to the right of the harbor is to the front of the warp bubble, and the constant portion to the left of the harbor is the back of the warp bubble.} \label{fig:4.3}
    \end{figure}

\subsection{Invariant Plots While Varying the Radius of the Warp Bubble}
    
\begin{figure}[ht]%
	\begin{subfigure}{.45\linewidth}
	    \centering
		\includegraphics[scale=0.25]{Images/R/Rs50000p100a1t1.png}
		\caption{Ricci scalar with $\rho=100$ m}
		\label{Rs50000p100a1t1 r}
	\end{subfigure}
	~
	\begin{subfigure}{.45\linewidth}
    	\centering
		\includegraphics[scale=0.25]{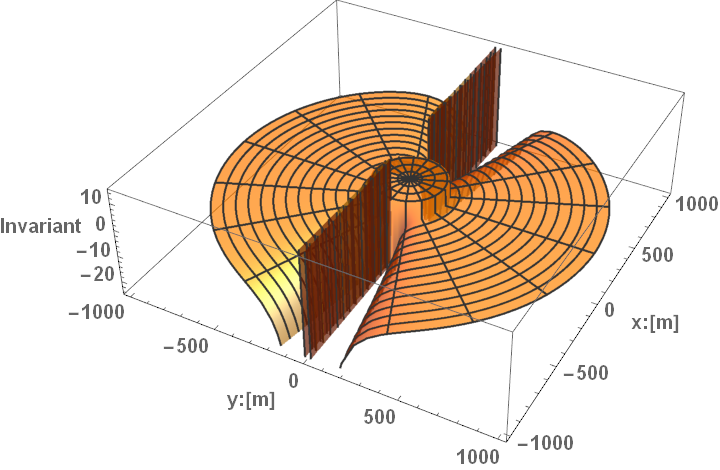}
		\caption{Ricci scalar with $\rho=200$ m}
		\label{Rs50000p200ap1t1 r}
	\end{subfigure}
	~
	\caption{The warp bubble radius, $\rho$, for the Ricci Scalar $R$ is plotted in the figures above.
	The other parameters were set to be $\sigma=50,000 \ \text{m}^{-1}$, $a=1.0 \ \text{m s}^{-2}$, and $t=1$ s. 
	The plots were rendered using Mathematica's\textsuperscript{\textregistered} RevolutionPlot3D.
	It plots a function by rotating the function around the z-axis.
	The amount that the space diverges from being flat is represented by the invariant function's magnitude, and its amount is labeled on the vertical axis.
	The x and y-axis are displayed on the plots, and they display the distance from a spaceship in the flat portion at the center of each figure.
    Each~radial line in the left hand column corresponds to a distance of approximately $33$ m and each radial line in the right hand column corresponds to a distance of approximately $67$ m.
    The~constant portion to the right of the harbor is to the front of the warp bubble, and the constant portion to the left of the harbor is the back of the warp bubble.} \label{fig:4.4}
    \end{figure}

    Varying the radius of the warp bubble, $\rho$, increases the size of the safe harbor inside each of the invariants.  
    Figure~\ref{fig:4.4} above presents the plots for each of the four invariants while doubling the skin depth and setting $\sigma=50,000 \ \text{m}^{-1}$, $a=1.0 \ \text{m s}^{-2}$, and $t=1$ s. 
    Figure~\ref{fig:9} in Appendix~\ref{app:fig} present the remaining invariants $r_1$, $r_2$, and $w_2$, which have similar features to the Ricci Scalar.
    In the figures for each invariant, the~radial coordinate doubles in radial size without affecting the shape of the plots. 
    The~safe harbor of $\rho\leq100$ m in the left hand column also doubles in size to $\rho\leq200$ m. 
    The~only other pertinent feature is in the internal structure of $w_2$. The structures are reduced implying that they cluster near the center. 
    
\section{Conclusions}
    This paper demonstrates how computing and plotting curvature invariants for various parameters of warp drive spacetimes can reveal their underlying features.
    While the individual functions require pages to express, their plots can be easily read and understood.
    The~plots give the magnitude of curvature at each point around the ship. 
    Where the plots' magnitudes are large, space is greatly warped and vice versa.
    Additionally, the~rate at which spacetime is being folded may be approximated by observing the changes in slopes on the plots.
    This information can be used to map the spacetime around the ship and potentially aid in navigation. 

    In this research, the~accelerating Nat\'ario warp drive metric was plotted for different choices of the parameters $t,a,\rho, \sigma$.
    In all plots of the listed parameters, the~constant features include a safe harbor and a wake running perpendicular to the warp bubble's direction of motion.
     The~wake will carry the spaceship along as Nat\'ario predicted.
    The~Ricci scalar revealed constant curvature outside of the harbor and wake.
    The~other invariants had asymptotically flat space outside of the harbor and wake.
    As a consequence, the~Ricci Scalar will have the greatest impact of the invariants on the spacetime outside of the harbor and wake. 
    By varying time, the~plot of each invariant experiences a sudden jump from positive curvature in the direction of motion to negative curvature. 
    As time progresses, the~shape of the $R$, $r_1$, and $r_2$ invariants remains constant, but the magnitude of the invariants increases.
    Also,~the~angular arc widens slightly for the invariants $r_1$ and $r_2$ as time increases. 
    After $100$ s, the~invariant $w_2$ begins to exhibit crenulations in the interior of the wake.
    Because the greatest invariant magnitudes occur for the invariant $w_2$, it must have the greatest impact on the wake and the internal structures.
    By varying the acceleration, the~invariant plots skip through the time slices and the internal structures become more prominent.
    Changing the skin depth did not change either the shape or magnitude of the invariant plots.
    As expected, doubling the radius also doubled the size of the safe harbor in the invariants plots without affecting the shape of the warp bubble.
    The~invariant plots give a rich and detailed understanding of the warp bubble's curvature.
    
    Computing and plotting the invariant functions have significant advantages for the inspection of warp drives and their potential navigation.
    As mentioned previously, plotting the invariants has the advantage that they are free from coordinate mapping distortions, divergences, discontinuities or other artifacts of the chosen coordinates.
    Once the plots of the invariant functions reveal the location of any artifacts, their position can be related mathematically to the standard tensors, and their effect on an objects motion can then be analyzed. 
    The~invariant plots properly illustrate the entire underlying spacetime independent of a chosen coordinate system. 
    A second advantage is the relative ease with which the invariants can be plotted. 
    Software packages exist or can be developed to calculate the standard tensors. 
    The~aforementioned tensors lead to a chosen basis of invariants. 
    While the CM invariants were chosen in this paper, other sets of invariants exist, such as the Cartan invariants and the Witten and Petrov invariants~\cite{16,8}.
    It is an open problem to inspect the curvature of the warp drive spacetimes in these invariant sets.
    It is expected that the main features identified in this paper will also hold in these different bases.  

    In addition to inspecting different invariant bases, further work can be done in mapping warp drive spacetimes.
    The~work in this paper can be further expanded to greater time slices for each invariant.
    Potentially, crenulations like the ones observed for $w_2$ also exist within the internal structures observed in the wakes of the other invariants.
    As the crenulations ripple, navigating the warp drive will be increasingly difficult.
    To minimize the effect of the crenulations, the~warp drive should be periodically turned on and off as it accelerates. 
    It can be speculated that the increase in the magnitude of the curvature as time advances establishes a relationship between the magnitude of the acceleration and the amount of energy required to accelerate the warp bubble to arbitrarily high velocities.
    Since~the acceleration skips through the warp drive's time slices, greater values of the acceleration will reach high magnitudes of each invariant and curvature more quickly.
    This relationship suggests that a realistic warp drive would only be able to accelerate to some finite velocity that is potentially greater than $c$, and it is evidence for a less restrictive superluminal censorship theorem than previously considered in~\cite{Barcelo:2002}.
    Instead, it is in accordance with the constraint that the net total energy stored in the warp bubble must be less than the total rest energy of the spaceship itself~\cite{LoboVisser}.
    In addition, the~crenulations imply that high frequency gravitational waves would be produced by an accelerating warp drive.
    Potentially, these waves could be detected by an extremely sensitive detector.
    A detector like the one proposed by~\cite{Woods} would be suitable.
    Another potential branch of the research presented in this paper is to calculate the geodesic equation for the accelerating Nat\'ario warp drive.
    The~Christoffel symbols $\Gamma^i_{jk}$ were calculated in the process to find the curvature invariants.
    It remains to plug them into the geodesic equation and solve given the initial conditions of being near Earth and at rest.
    Finally, the~technique of plotting the invariants can be applied to other warp drive spacetimes such as Alcubierre, Krasnikov, Van Den Broeck, and Nat\'ario at a constant velocity.

\newpage

\section{Acknowledgements}
The authors would like to thank F.~Loup and D.~D.~McNutt for beneficial discussions.	

\appendix
\section{Null Vectors of the Nat\'ario Metric} \label{app:null}
    A null tetrad contains two real null vectors, $\bf{k}$ and $\bf{l}$, and two complex conjugate null vectors, $\bf{m}$ and $\bf{\bar{m}}$ that satisfy the following algebraic relationships 
    \cite{Stephani:2003tm}:
    \begin{align}
        \bf{e}_a &=(\bf{l}, \bf{k}, \bf{m},\bf{\bar{m}}) \label{21}, \\
        g_{ab} &=2m_{(a}\bar{m}_{b)}-2k_{(a}l_{b)}=\begin{pmatrix}0 & 1 & 0 & 0 \\ 1 & 0 & 0 & 0 \\ 0 & 0 & 0 & -1 \\ 0 & 0 & -1 & 0 \end{pmatrix}. \label{22}
    \end{align}
    Given an orthonormal tetrad, $\bf{E}_a$, it can be related to a complex null tetrad Equation \eqref{21} by:
    \begin{align}
        l_a &=\frac{1}{\sqrt{2}}(E_1 + E_2), &
        k_a &=\frac{1}{\sqrt{2}}(E_1 - E_2), \nonumber\\
        m_a &=\frac{1}{\sqrt{2}}(E_3 + i E_4), & 
        \bar{m}_a &=\frac{1}{\sqrt{2}}(E_3 - i E_4). \label{23}
    \end{align}
    The line element in Equation \eqref{eq:metric} has an orthonormal tetrad:
    \begin{align}
        E_1&=\begin{pmatrix}1 - X_t \ 0 \ 0 \ 0\end{pmatrix}, &
        E_2&=\begin{pmatrix}X_{r_s} \ -1 \ 0 \ 0\end{pmatrix}, \nonumber \\  E_3&=\begin{pmatrix}X_{\theta} \ 0 \ -r \ 0\end{pmatrix}, & E_4&=\begin{pmatrix}0 \ 0 \ 0 \ r \sin{\theta}\end{pmatrix}.  \label{24}
    \end{align}
    Using Mathematica\textsuperscript{\textregistered}, \eqref{22} and \eqref{24} satisfy the relationship 
    \begin{equation}
        g_{ab}=E_a\cdot E_b. \label{25}
    \end{equation}
    By applying Eqs.~\eqref{23} to \eqref{24}, the null vectors in Eq.~\eqref{eq:null} result. 
\newpage 

\section{Invariant Plots} \label{app:fig}
\renewcommand{\thefigure}{A\arabic{figure}}
\setcounter{figure}{0}
    
    \begin{figure}[ht]
	\begin{subfigure}[t]{.48\linewidth}
	    \centering
		\includegraphics[scale=0.25]{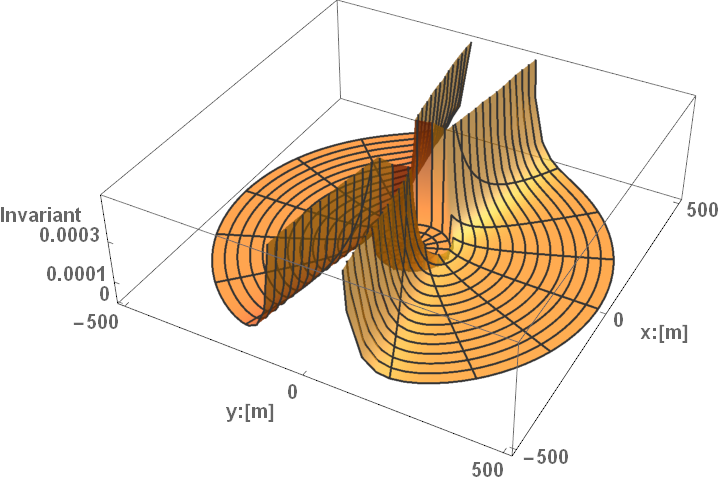}
		\caption{$r_1$ and $t=0.0$ s}
		\label{r1s50000p100a1t0}
	\end{subfigure}
	~
	\begin{subfigure}[t]{.48\linewidth}
	    \centering
		\includegraphics[scale=0.25]{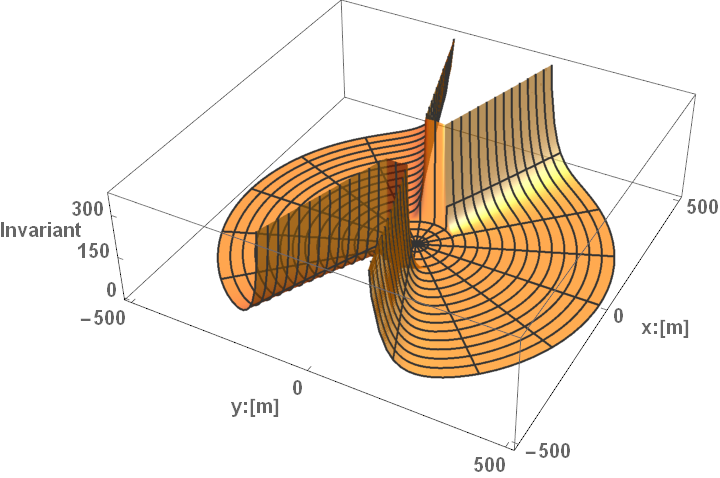}
		\caption{$r_1$ and $t=1$ s}
		\label{r1s50000p100a1t1}
	\end{subfigure}
	\par\bigskip
	\begin{subfigure}[t]{.48\linewidth}
	    \centering
		\includegraphics[scale=0.25]{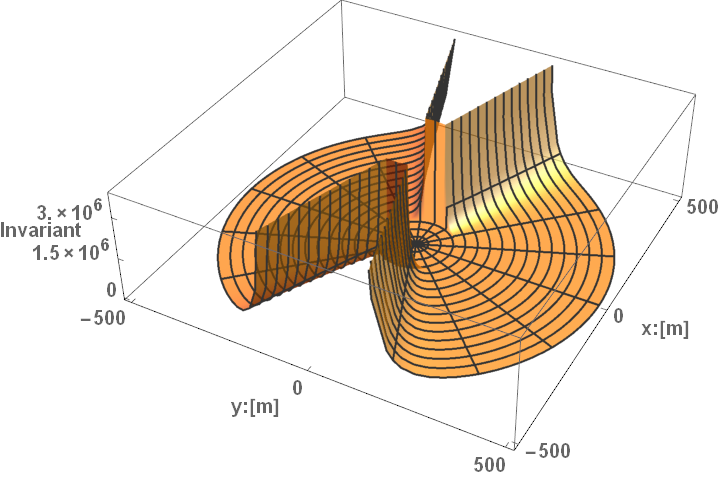}
		\caption{$r_1$ and $t=10$ s}
		\label{r1s50000p100a1t10}
	\end{subfigure}
	~
	\begin{subfigure}[t]{.5\linewidth}
	    \centering
		\includegraphics[scale=0.25]{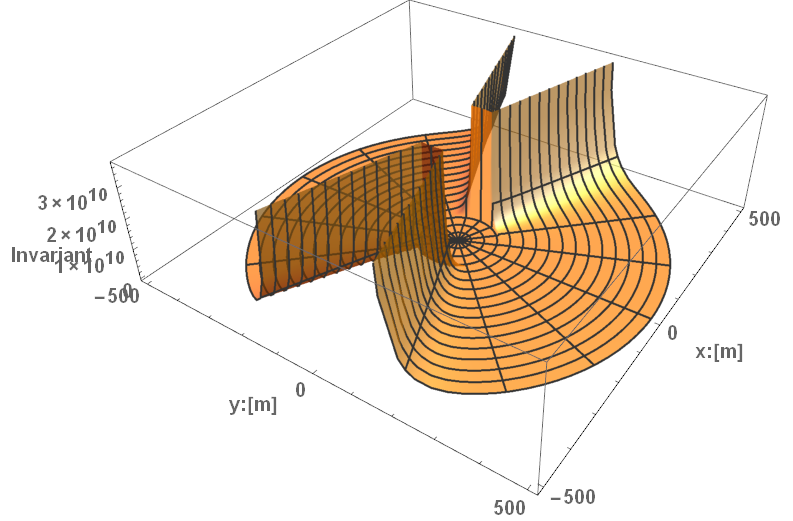}
		\caption{$r_1$ and $t=100$ s}
		\label{r1s50000p100a1t100}
	\end{subfigure}
	~
    \vspace{.25cm}
	~
	\caption{The time, $t$, of the invariant $r_1$ is plotted in the figures above.   
	The other parameters were set to be $\sigma=50,000 \ \text{m}^{-1}$, $\rho=100$ m, and $a=1.0 \ \text{m s}^{-2}$.
	The plots were rendered using Mathematica's\textsuperscript{\textregistered} RevolutionPlot3D.
	It plots a function by rotating the function around the z-axis.
	The amount that the space diverges from being flat is represented by the invariant function's magnitude, and its amount is labeled on the vertical axis.
	The x and y-axis are displayed on the plots, and they display the distance from a spaceship in the flat portion at the center of each figure.
    Each radial line in the plot corresponds to a distance of approximately $33$ m.
    The~constant portion to the right of the harbor is to the front of the warp bubble, and the constant portion to the left of the harbor is the back of the warp bubble.} \label{fig:2}
    \end{figure}
    
    
    \begin{figure}[ht]
	\begin{subfigure}[t]{.48\linewidth}
	    \centering
		\includegraphics[scale=0.25]{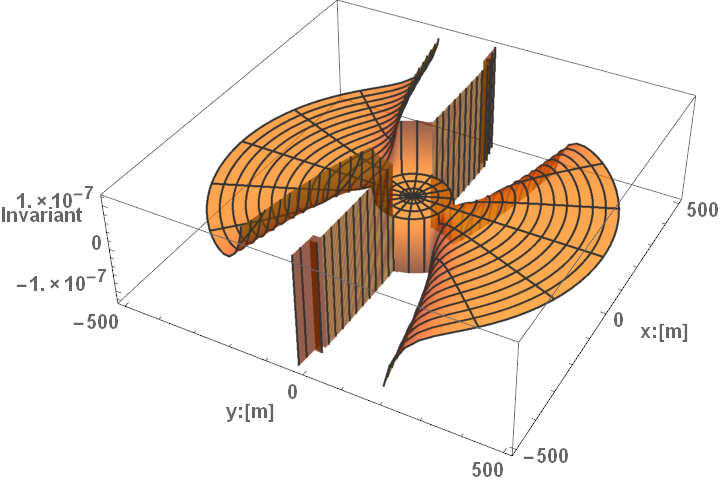}
		\caption{$r_2$ and $t=0.0$ s}
		\label{r2s50000p100a1t0}
	\end{subfigure}
	~
	\begin{subfigure}[t]{.48\linewidth}
	    \centering
		\includegraphics[scale=0.25]{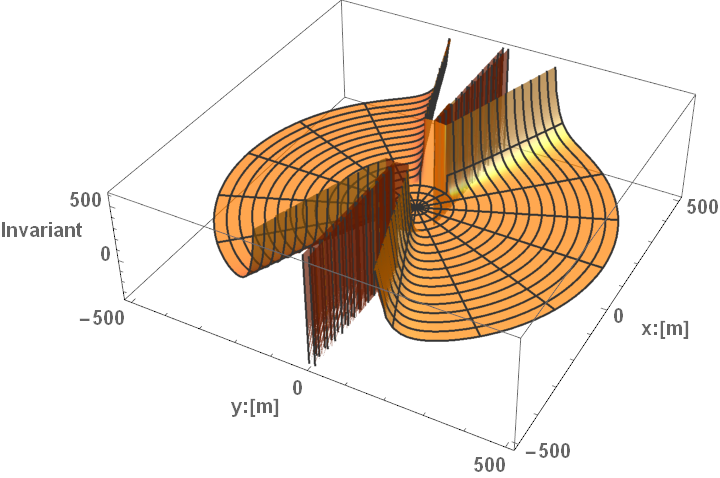}
		\caption{$r_2$ and $t=1$ s}
		\label{r2s50000p100a1t1}
	\end{subfigure}
	\caption{\textit{Cont.}}
        \label{fig:a2b}
	\end{figure}

	\begin{figure}[ht]\ContinuedFloat
	\centering
	\begin{subfigure}[t]{.48\linewidth}
	    \centering
		\includegraphics[scale=0.25]{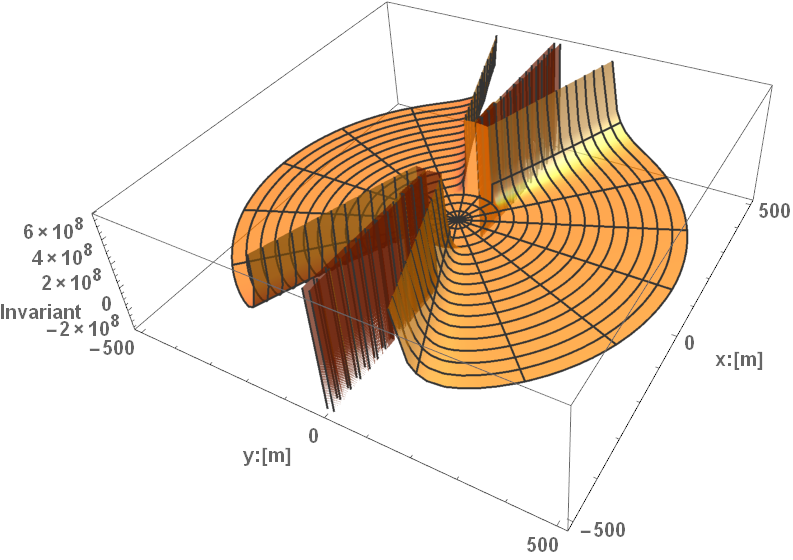}
		\caption{$r_2$ and $t=10$ s}
		\label{r2s50000p100a1t10}
	\end{subfigure}
	~
	\begin{subfigure}[t]{.5\linewidth}
	    \centering
		\includegraphics[scale=0.25]{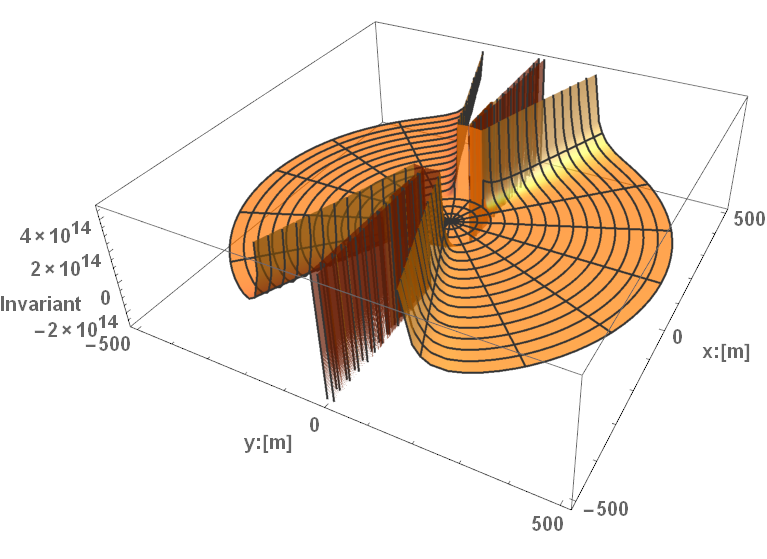}
		\caption{$r_2$ and $t=100$ s}
		\label{r2s50000p100a1t100}
	\end{subfigure}
	~
    \vspace{.25cm}
    ~
	\caption{The time, $t$, of the invariant $r_2$ is plotted in the figures above.  
	The other parameters were set to be $\sigma=50,000 \ \text{m}^{-1}$, $\rho=100$ m, and $a= 1.0 \ \text{m s}^{-2}$. 
	The plots were rendered using Mathematica's\textsuperscript{\textregistered} RevolutionPlot3D.
	It plots a function by rotating the function around the z-axis.
	The amount that the space diverges from being flat is represented by the invariant function's magnitude, and its amount is labeled on the vertical axis.
	The x and y-axis are displayed on the plots, and they display the distance from a spaceship in the flat portion at the center of each figure.
    Each radial line in the plot corresponds to a distance of approximately $33$ m.
    The~constant portion to the right of the harbor is to the front of the warp bubble, and the constant portion to the left of the harbor is the back of the warp bubble.} \label{fig:3}
    \end{figure}
    \unskip
    \begin{figure}[ht]
	\begin{subfigure}[t]{.48\linewidth}
	    \centering
		\includegraphics[scale=0.25]{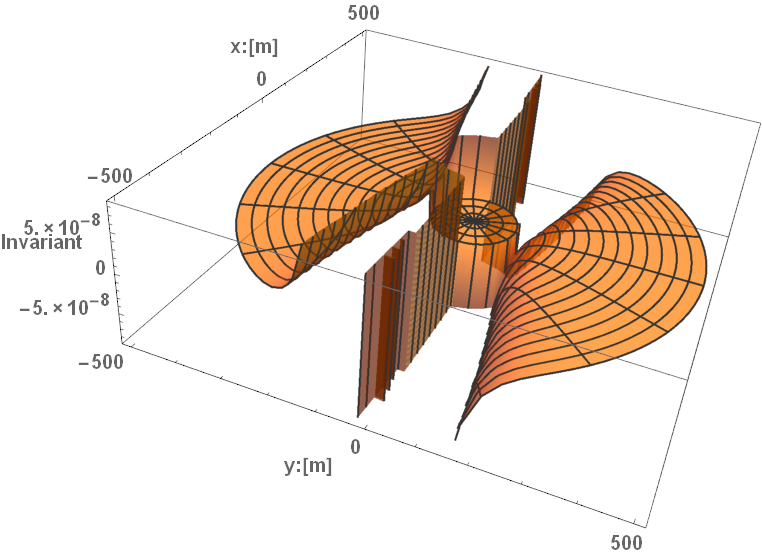}
		\caption{$w_2$ and $t=0$ s}
		\label{w2s50000p100a1t0}
	\end{subfigure}
	~
	\begin{subfigure}[t]{.48\linewidth}
	    \centering
		\includegraphics[scale=0.25]{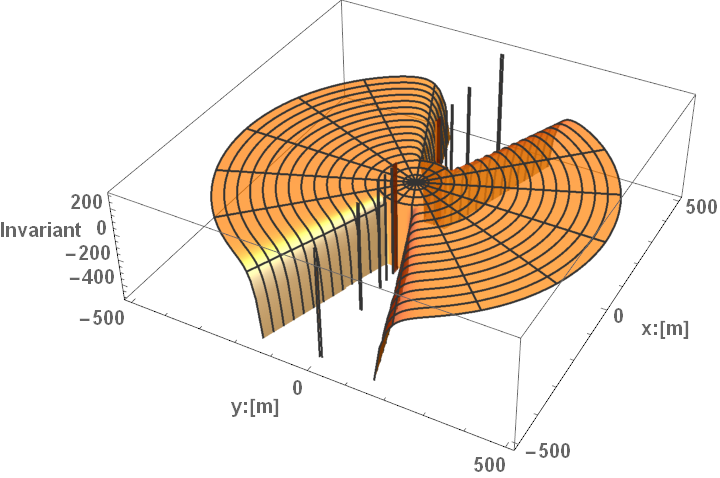}
		\caption{$w_2$ and $t=1.0$ s}
		\label{w2s50000p100a1t1 t}
	\end{subfigure}
	\par\bigskip
	\begin{subfigure}[t]{.48\linewidth}
	    \centering
		\includegraphics[scale=0.25]{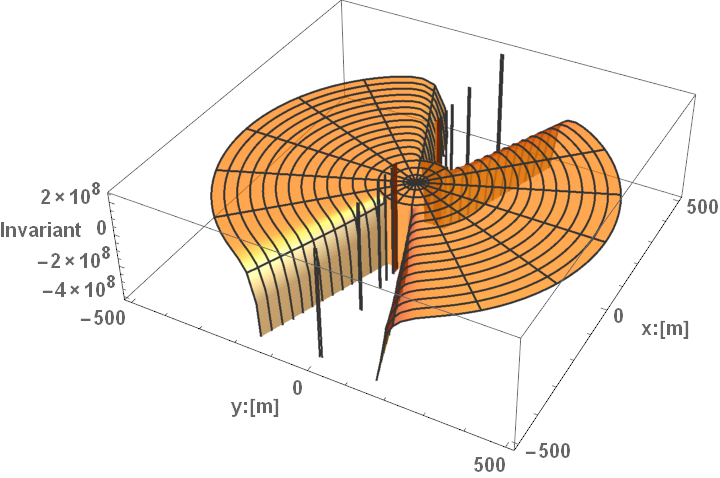}
		\caption{$w_2$ and $t=10.0$ s}
		\label{w2s50000p100a1t10}
	\end{subfigure}
	~
	\begin{subfigure}[t]{.48\linewidth}
	    \centering
		\includegraphics[scale=0.25]{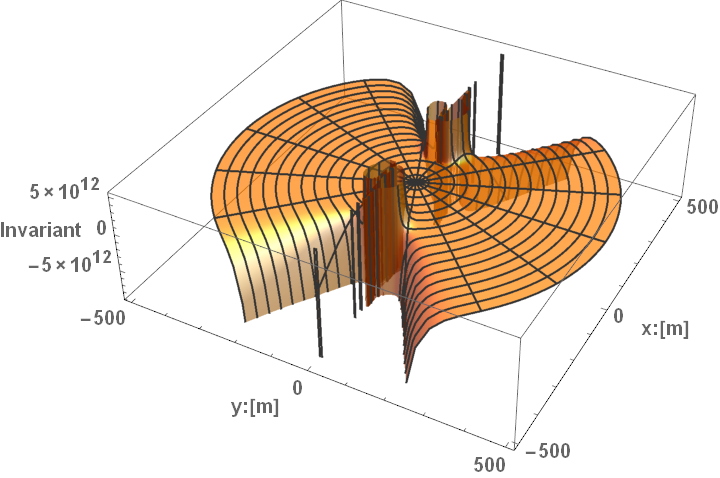}
		\caption{$w_2$ and $t=100.0$ s}
		\label{w2s50000p100a1t100}
	\end{subfigure}
	\caption{\textit{Cont.}}
        \label{fig:a3d}
	\end{figure}
	
	\begin{figure}[ht]\ContinuedFloat
	    \centering
	\begin{subfigure}[t]{.5\linewidth}
	    \centering
		\includegraphics[scale=0.25]{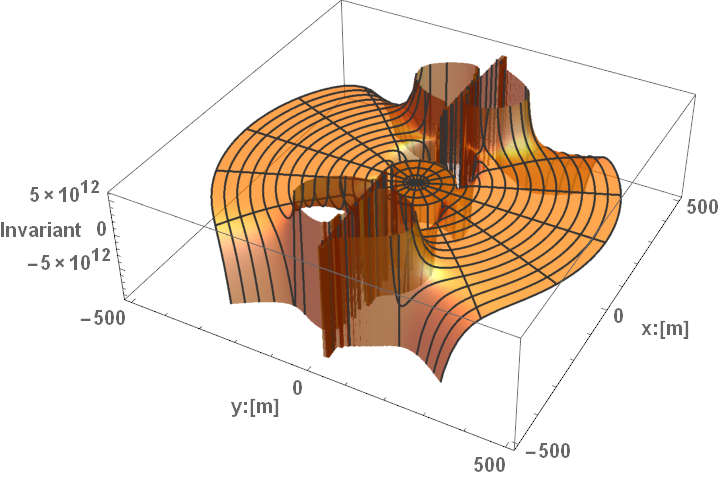}
		\caption{$w_2$ and $t=200.0$ s}
		\label{w2s50000p100a1t200}
	\end{subfigure}
	~
	\begin{subfigure}[t]{.45\linewidth}
	    \centering
		\includegraphics[scale=0.25]{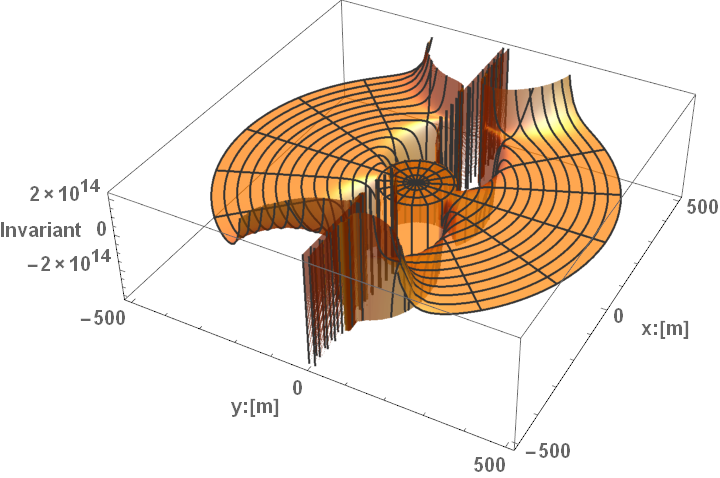}
		\caption{$w_2$ and $t=300.0$ s}
		\label{w2s50000p100a1t300}
	\end{subfigure}
	~
    \vspace{.25cm}
    ~
    \caption{The time, $t$, of the invariant $w_2$ is plotted in the figures above. 
    The~other parameters were set to be $\sigma=50,000 \ \text{m}^{-1}$, $\rho=100$ m, and $a= 1.0 \ \text{m s}^{-2}$. 
    The~plots were rendered using Mathematica's\textsuperscript{\textregistered} RevolutionPlot3D.
	It plots a function by rotating the function around the z-axis.
	The amount that the space diverges from being flat is represented by the invariant function's magnitude, and its amount is labeled on the vertical axis.
	The x and y-axis are displayed on the plots, and they display the distance from a spaceship in the flat portion at the center of each figure.
    Each radial line in the plot corresponds to a distance of approximately $33$ m.
    The~constant portion to the right of the harbor is to the front of the warp bubble, and the constant portion to the left of the harbor is the back of the warp bubble.} \label{fig:4}
    \end{figure}
    
    
    \begin{figure}[ht]
	\begin{subfigure}[t]{.47\linewidth}
	    \centering
		\includegraphics[scale=0.25]{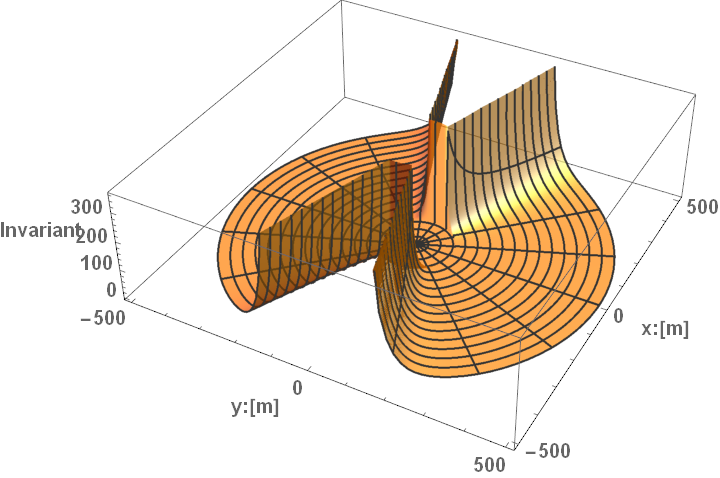}
		\caption{$r_1$ and $a=0.1 \ \text{m s}^{-2}$}
		\label{r1s50000p100ap1t1}
	\end{subfigure}
	~
	\begin{subfigure}[t]{.47\linewidth}
	    \centering
		\includegraphics[scale=0.25]{Images/r1/r1s50000p100a1t1.png}
		\caption{$r_1$ and $a=1.0 \ \text{m s}^{-2}$}
		\label{r1s50000p100a1t1 2}
	\end{subfigure}
	\par\bigskip
	\begin{subfigure}[t]{1.0\linewidth}
	    \centering
		\includegraphics[scale=0.25]{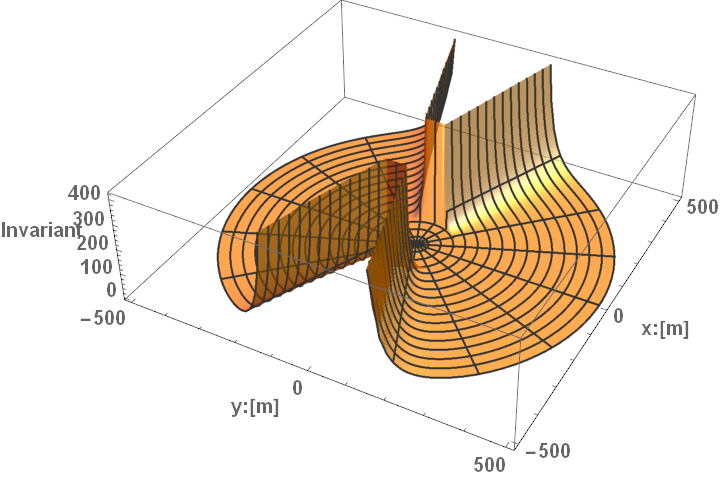}
		\caption{$r_1$ and $a=10.0 \ \text{m s}^{-2}$}
		\label{r1s50000p100a10t1}
	\end{subfigure}
	~
    \vspace{.25cm}
    ~
	\caption{The acceleration, $a$, of the invariant $r_1$ is plotted in the figures above. 
	The other parameters were set to be $\sigma=50,000 \ \text{m}^{-1}$, $\rho=100$ m, and $t=1$ s.
	The plots were rendered using Mathematica's\textsuperscript{\textregistered} RevolutionPlot3D.
	It plots a function by rotating the function around the z-axis.
	The amount that the space diverges from being flat is represented by the invariant function's magnitude, and its amount is labeled on the vertical axis.
	The x and y-axis are displayed on the plots, and they display the distance from a spaceship in the flat portion at the center of each figure.
    Each radial line in the plot corresponds to a distance of approximately $33$ m.
    The~constant portion to the right of the harbor is to the front of the warp bubble, and the constant portion to the left of the harbor is the back of the warp bubble.} \label{fig:6}
    \end{figure}
    
    
	\begin{figure}[ht]
	\begin{subfigure}[t]{.47\linewidth}
	    \centering
		\includegraphics[scale=0.25]{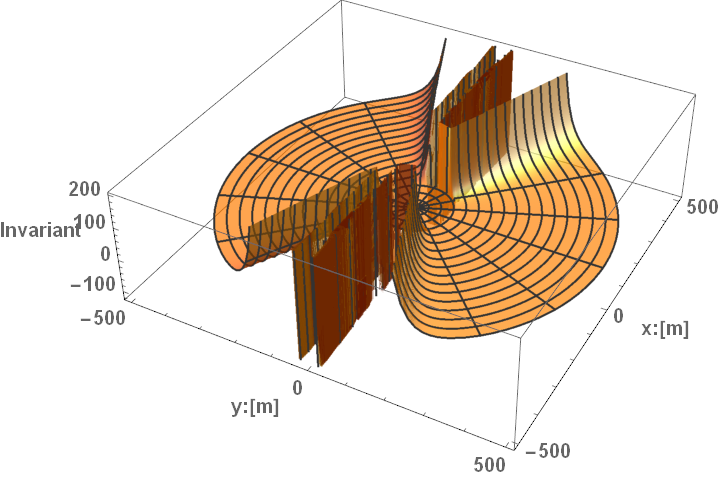}
		\caption{$r_2$ and $a=0.1 \ \text{m s}^{-2}$}
		\label{r2s50000p100ap1t1}
	\end{subfigure}
	~
	\begin{subfigure}[t]{.45\linewidth}
	    \centering
		\includegraphics[scale=0.25]{Images/r2/r2s50000p100a1t1.png}
		\caption{$r_2$ and $a=1.0 \ \text{m s}^{-2}$}
		\label{r2s50000p100a1t1 2}
	\end{subfigure}
	\par\bigskip
	\begin{subfigure}[t]{1.0\linewidth}
	    \centering
		\includegraphics[scale=0.25]{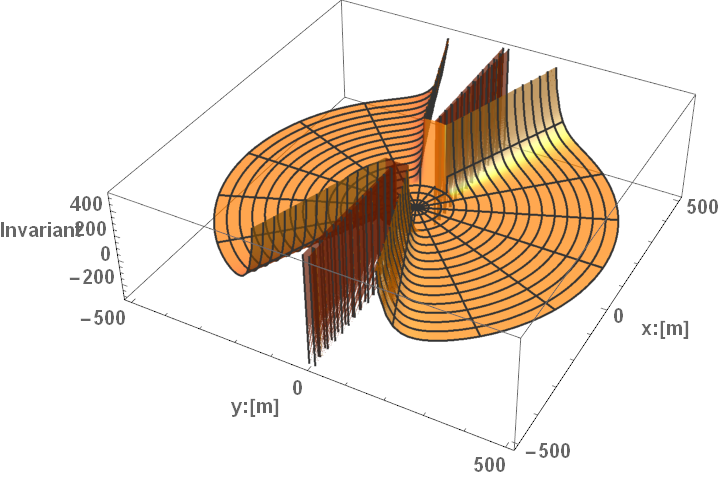}
		\caption{$r_2$ and $a=10.0 \ \text{m s}^{-2}$}
		\label{r2s50000p100a10t1}
	\end{subfigure}
	~
    \vspace{.25cm}
    ~
    \caption{The acceleration, $a$, of the invariant $r_2$ is plotted in the figures above. 
    The~other parameters were set to be $\sigma=50,000 \ \text{m}^{-1}$, $\rho=100$ m, and $t=1$ s.
    The~plots were rendered using Mathematica's\textsuperscript{\textregistered} RevolutionPlot3D.
	It plots a function by rotating the function around the z-axis.
	The amount that the space diverges from being flat is represented by the invariant function's magnitude, and its amount is labeled on the vertical axis.
	The x and y-axis are displayed on the plots, and they display the distance from a spaceship in the flat portion at the center of each figure.
    Each radial line in the plot corresponds to a distance of approximately $33$ m.
    The~constant portion to the right of the harbor is to the front of the warp bubble, and the constant portion to the left of the harbor is the back of the warp bubble.} \label{fig:7}
    \end{figure}
    
    
    \begin{figure}[ht]
	\begin{subfigure}[t]{.45\linewidth}
	    \centering
		\includegraphics[scale=0.25]{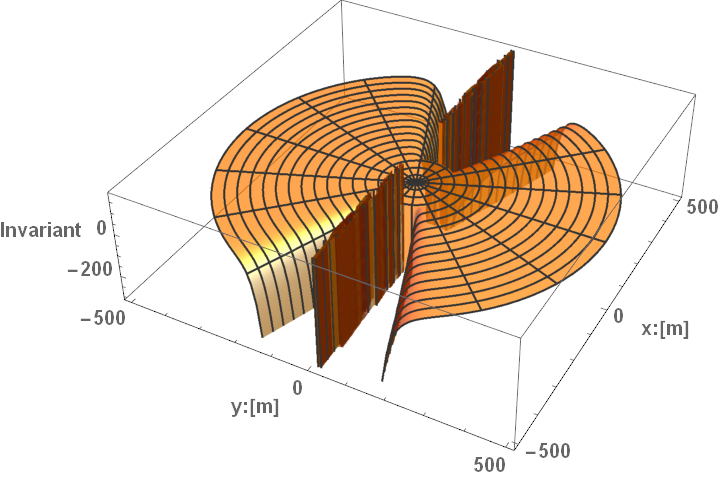}
		\caption{$w_2$ and $a=0.1 \ \text{m s}^{-2}$}
		\label{w2s50000p100ap1t1}
	\end{subfigure}
	~
	\begin{subfigure}[t]{.45\linewidth}	
	\centering	
	    \includegraphics[scale=0.25]{Images/w2/w2s50000p100a1t1.png}
		\caption{$w_2$ and $a=1.0 \ \text{m s}^{-2}$}
		\label{w2s50000p100a1t1 2}
	\end{subfigure}
	\caption{\textit{Cont.}}
        \label{fig:a6b}
	\end{figure}
	
	\begin{figure}[ht]\ContinuedFloat
	    \centering
	\begin{subfigure}[t]{1.0\linewidth}
	    \centering
		\includegraphics[scale=0.25]{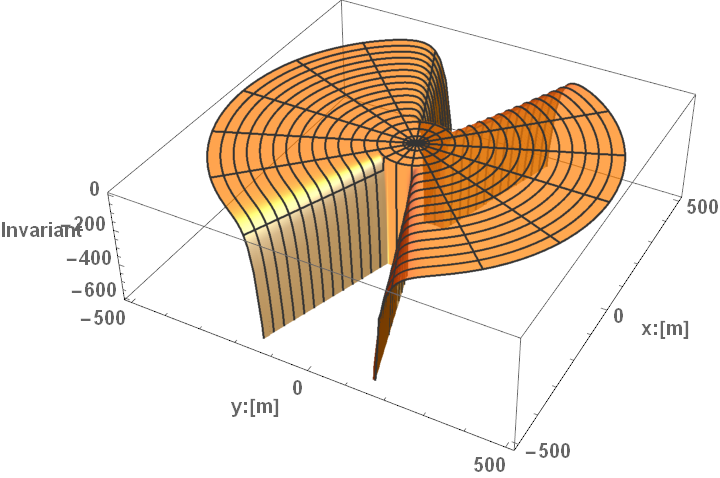}
		\caption{$w_2$ and $a=10.0 \ \text{m s}^{-2}$}
		\label{w2s50000p100a10t1}
	\end{subfigure}
	~
    \vspace{.25cm}
    ~
    \caption{The acceleration, $a$, of the invariant $w_2$ is plotted in the figures above. 
    The~other parameters were set to be $\sigma=50,000 \ \text{m}^{-1}$, $\rho=100$ m, and $t=1$ s.
    The~plots were rendered using Mathematica's\textsuperscript{\textregistered} RevolutionPlot3D.
	It plots a function by rotating the function around the z-axis.
	The amount that the space diverges from being flat is represented by the invariant function's magnitude, and its amount is labeled on the vertical axis.
	The x and y-axis are displayed on the plots, and they display the distance from a spaceship in the flat portion at the center of each figure.
    Each radial line in the plot corresponds to a distance of approximately $33$ m.
    The~constant portion to the right of the harbor is to the front of the warp bubble, and the constant portion to the left of the harbor is the back of the warp bubble.} \label{fig:8}
    \end{figure}
    ~
    \begin{figure}[ht]%
	\begin{subfigure}{.5\linewidth}
	\centering
		\includegraphics[scale=0.25]{Images/r1/r1s50000p100a1t1.png}
		\caption{The invariant $r_1$ with $\sigma=50,000 \text{m}^{-1}$}
		\label{r1s50000p100a1t1 s}
	\end{subfigure}	
	\begin{subfigure}{.5\linewidth}
	    \centering
		\includegraphics[scale=0.25]{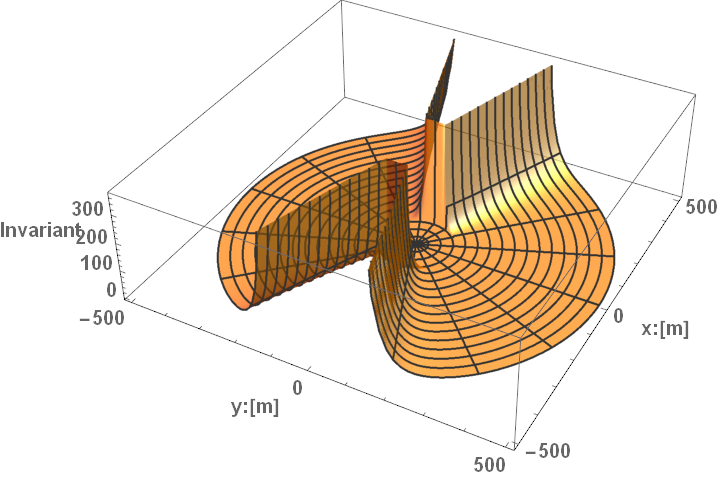}
		\caption{The invariant $r_1$ with $\sigma=100,000 \text{m}^{-1}$}
		\label{r1s100000p200a1t1 s}
	\end{subfigure}
	\par\bigskip
	\begin{subfigure}{.5\linewidth}
	    \centering
		\includegraphics[scale=0.25]{Images/r2/r2s50000p100a1t1.png}
		\caption{The invariant $r_2$ with $\sigma=50,000 \ \text{m}^{-1}$}
		\label{r2s50000p100a1t1 s}
	\end{subfigure}
	~
	\begin{subfigure}{.5\linewidth}
	    \centering
		\includegraphics[scale=0.25]{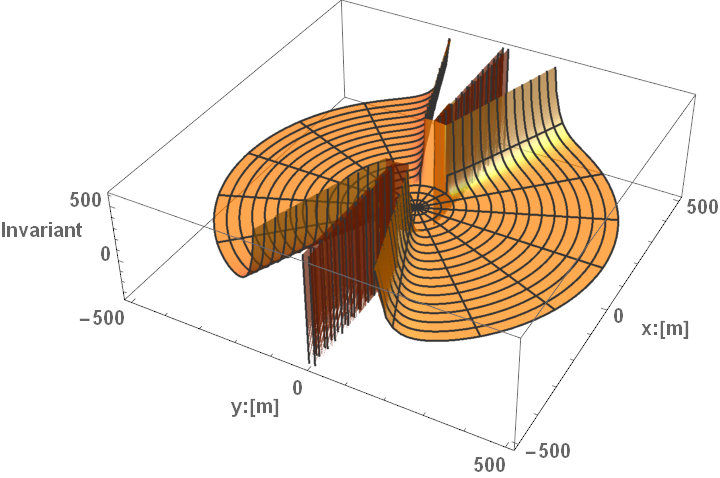}
		\caption{The invariant $r_2$ with $\sigma=100,000 \ \text{m}^{-1}$}
		\label{r2s50000p200a1t1 s}
	\end{subfigure}
      \caption{\textit{Cont.}}
        \label{fig:a7d}
	\end{figure}
	
	\begin{figure}[ht]\ContinuedFloat
	    \centering
	\begin{subfigure}{.45\linewidth}
		\includegraphics[scale=0.25]{Images/w2/w2s50000p100a1t1.png}
		\caption{The invariant $w_2$ with $\sigma=50,000 \text{m}^{-1}$}
		\label{w2s50000p100a1t1 s}
	\end{subfigure}
	~
	\begin{subfigure}{0.45\linewidth}
	    \centering
		\includegraphics[scale=0.25]{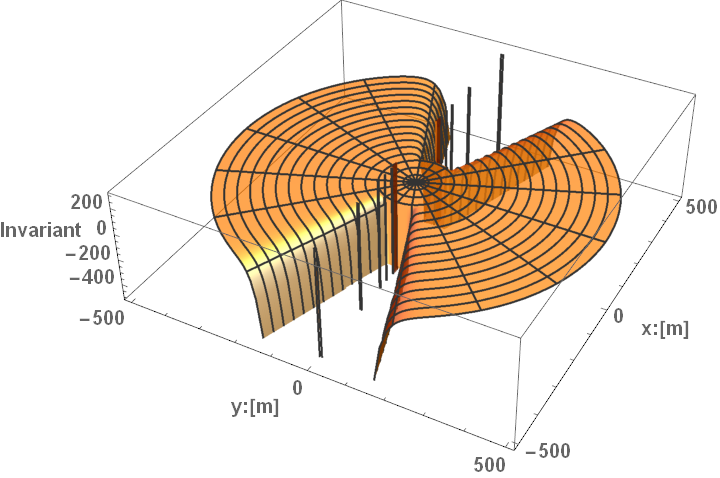}
		\caption{The invariant $w_2$ with $\sigma=100,000 \text{m}^{-1}$}
		\label{w2s50000p200a1t1 s}
	\end{subfigure}
	~
	\vspace{.25cm}
	~
	\caption{The warp bubble skin depth, $\sigma$, of the invariants $r_1$, $r_2$, and $w_2$ are plotted in the figures above.
	The other parameters were set to be $\rho=100$ m, $a=1.0 \ \text{m s}^{-2}$, and $t=1$ s. 
	The plots were rendered using Mathematica's\textsuperscript{\textregistered} RevolutionPlot3D.
	It plots a function by rotating the function around the z-axis.
	The amount that the space diverges from being flat is represented by the invariant function's magnitude, and its amount is labeled on the vertical axis.
	The x and y-axis are displayed on the plots, and they display the distance from a spaceship in the flat portion at the center of each figure.
    Each~radial line in the left hand column corresponds to a distance of approximately $33$ m.
    The~constant portion to the right of the harbor is to the front of the warp bubble, and the constant portion to the left of the harbor is the back of the warp bubble.} \label{fig:9}
    \end{figure}
    ~
    \begin{figure}[ht]%
	\begin{subfigure}{.45\linewidth}
	\centering
		\includegraphics[scale=0.25]{Images/r1/r1s50000p100a1t1.png}
		\caption{The invariant $r_1$ with $\rho=100$ m}
		\label{r1s50000p100a1t1 r}
	\end{subfigure}
	~
	\begin{subfigure}{.45\linewidth}
	    \centering
		\includegraphics[scale=0.25]{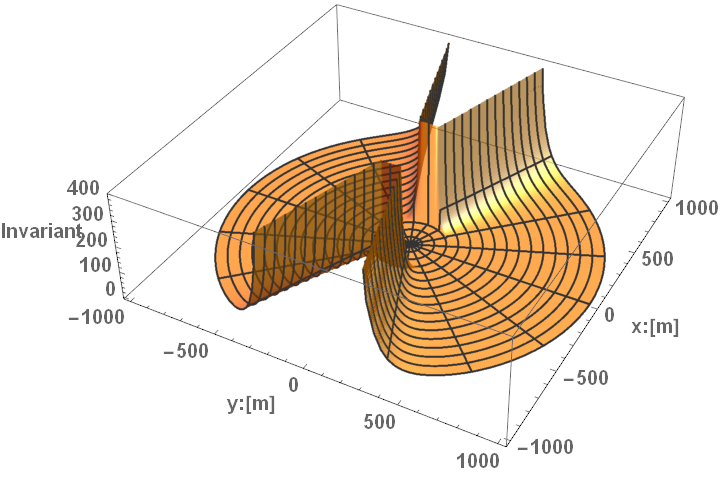}
		\caption{The invariant $r_1$ with $\rho=200$ m}
		\label{r1s50000p200a1t1 r}
	\end{subfigure}
	\par\bigskip
	\begin{subfigure}{.45\linewidth}
	    \centering
		\includegraphics[scale=0.25]{Images/r2/r2s50000p100a1t1.png}
		\caption{The invariant $r_2$ with $\rho=100$ m}
		\label{r2s50000p100a1t1 r}
	\end{subfigure}
	~
	\begin{subfigure}{.45\linewidth}
	    \centering
		\includegraphics[scale=0.25]{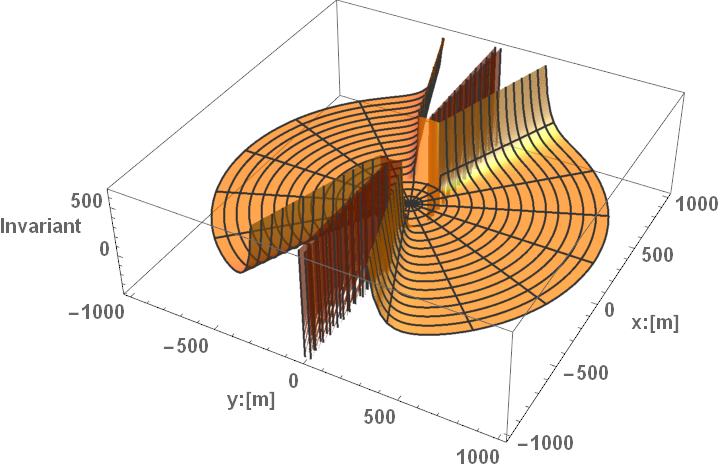}
		\caption{The invariant $r_2$ with $\rho=200 $ m}
		\label{r2s50000p200a1t1 r}
	\end{subfigure}
	\caption{\textit{Cont.}}
        \label{fig:a8d}
	\end{figure}
	
	\begin{figure}[ht]\ContinuedFloat
	    \centering
\par\bigskip
	\begin{subfigure}{.45\linewidth}
	\centering
		\includegraphics[scale=0.25]{Images/w2/w2s50000p100a1t1.png}
		\caption{The invariant $w_2$ with $\rho=100$ m}
		\label{w2s50000p100a1t1 r}
	\end{subfigure}
	~
	\begin{subfigure}{0.45\linewidth}
	    \centering
		\includegraphics[scale=0.25]{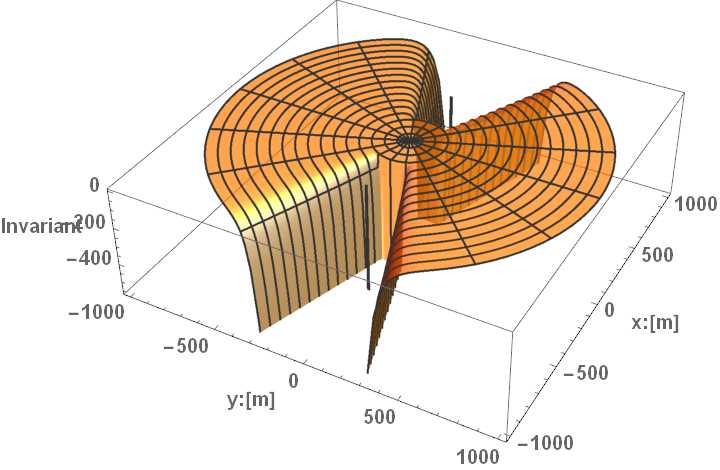}
		\caption{The invariant $w_2$ with $\rho=200m$}
		\label{w2s50000p200a1t1 r}
	\end{subfigure}
	~
	\vspace{.25cm}
	~
	\caption{The warp bubble radius, $\rho$, of the invariants $r_1$, $r_2$, and $w_2$ are plotted in the figures above.
	The other parameters were set to be $\sigma=50,000 \ \text{m}^{-1}$, $a=1.0 \ \text{m s}^{-2}$, and $t=1$ s. 
	The plots were rendered using Mathematica's\textsuperscript{\textregistered} RevolutionPlot3D.
	It plots a function by rotating the function around the z-axis.
	The amount that the space diverges from being flat is represented by the invariant function's magnitude, and its amount is labeled on the vertical axis.
	The x and y-axis are displayed on the plots, and they display the distance from a spaceship in the flat portion at the center of each figure.
    Each~radial line in the left hand column corresponds to a distance of approximately $33$ m and each radial line in the right hand column corresponds to a distance of approximately $67$ m.
    The~constant portion to the right of the harbor is to the front of the warp bubble, and the constant portion to the left of the harbor is the back of the warp bubble.} \label{fig:10}
    \end{figure}

\FloatBarrier


\begin{thebibliography}{1}
		
\bibitem{Alcubierre:1994tu} 
Alcubierre, M. The Warp drive: Hyperfast travel within general relativity. \emph{Class. Quantum Gravity} \textbf{1994}, \emph{{11}}, L73, doi:10.1088/0264-9381/11/5/001.
\bibitem{Davis}
Davis, E.W. Chapter 15: Faster-Than-Light Approaches in General Relativity. In \emph{Frontiers of Propulsion Science}; (2nd Printing with Corrections, 2012); Progress in Astronautics \& Aeronautics Series;   American Institution of Aeronautics \& Astronautics Press: Reston, VA, USA, 2009; Volume~227, pp.~473--509.
\bibitem{Natario:2001tk} 
Nat\'ario, J. 
Warp drive with zero expansion.
\emph{Class. Quantum Gravity} \textbf{2002}, \emph{{19}}, 1157, doi:10.1088/0264-9381/19/6/308.
\bibitem{Loup}
Loup, F. An Extended Version of the Nat\'ario Warp Drive Equation Based in the Original 3 + 1 ADM Formalism Which Encompasses Accelerations and Variable Velocities.  Ph.D. Thesis, Residencia de Estudantes Universitas Lisboa Portugal, $<hal-01655423>$, 2017. 
\bibitem{Loup2}
Loup, F. \emph{Six Different Natario Warp Drive Spacetime Metric Equations}; Research Report;
Residencia de Estudantes Universitas Lisboa Portugal. $<ffhal-01862911f.>$~{2018}. 
\bibitem{Krasnikov:1995ad} 
Krasnikov, S.V. Hyperfast travel in general relativity. \emph{Phys.~Rev.~D.} \textbf{1998}, \emph{{57}}, 4760, doi:10.1103/PhysRevD.57.4760.
\bibitem{VanDenBroeck:1999sn} 
Van Den Broeck, C. A 'Warp drive' with reasonable total energy requirements. \emph{Class. Quantum Gravity} \textbf{1999},~\emph{{16}}, 3973, doi:10.1088/0264-9381/16/12/314.
\bibitem{5} 
Morris, M.S.; Thorne, K.S.~
Wormholes in space-time and their use for interstellar travel: A tool for teaching general relativity.
\emph{Am.~J.~Phys.} \textbf{1988}, \emph{{56}}, 395, doi:10.1119/1.15620.
\bibitem{6} Morris, M.S.; Thorne, K.S.; Yurtsever, U. Wormholes, time machines, and the weak energy conditions. \emph{Phys.~Rev.~Lett.} \textbf{1988},~\emph{61},~1446--1449.
\bibitem{9} Visser, M. \emph{Lorentzian Wormholes: From Einstein to Hawking}; AIP Press: New York, NY, USA,~1995.
\bibitem{11} Lobo, F.S.N. Wormhole Basics. 
In \textit{Wormholes, Warp Drives and Energy Conditions}; Fund.~Theor.~Phys. \textbf{189}
 Lobo, F.S.N., Ed.; Springer: Cham, Switzerland, {2017; pp.~11--33}, doi:10.1007/978-3-319-55182-1
\bibitem{3} Mattingly, B.; Kar, A.; Julius, W.; Gorban, M.; Watson, C.; Ali, M.D.; Baas, A.; Elmore, C.; Shakerin, B.; Davis, E.W.; et al. 
Curvature Invariants for Lorentzian Traversable Wormholes.
\emph{Universe} \textbf{2020}, \emph{{6}}, 11, doi:10.3390/universe6010011.
\bibitem{Santos-Pereira:2020puq}
Santos-Pereira, O.L.; Abreu, E.M.C.; Ribeiro, M.B.
Dust content solutions for the Alcubierre warp drive spacetime.
\emph{Eur. Phys. J. C} \textbf{2020}, \emph{{80}}, 786, doi:10.1140/epjc/s10052-020-8355-2.
\bibitem{Chris} Christoffel, E.B. Ueber die Transformation der homogenen Differentialausdrücke zweiten Grades. \emph{J. Die Reine Angew. Math.} \textbf{1869},~\emph{70},~46--70.
\bibitem{ZM} Zakhary, E.; McIntosh, C.B.G. A Complete Set of Riemann Invariants. \emph{Gen.~Relat.~Gravit.} \textbf{1997},~\emph{29},~539--581.
\bibitem{CM} Carminati, J.; McLenaghan, R.G. Algebraic invariants of the Riemann tensor in a four‐dimensional Lorentzian space. \emph{J. Math. Phys.} \textbf{1991}, \emph{32},~3135--3140,
doi:10.1063/1.529470.
\bibitem{Santosuosso:1998he} 
Santosuosso, K.; Pollney, D.; Pelavas, N.; Musgrave, P.; Lake, K.
Invariants of the Riemann tensor for class B warped product spacetimes.
\emph{Comput. Phys. Commun.} \textbf{1998}, \emph{{115}}, 381,
doi:10.1016/S0010-4655(98)00134-9.
\bibitem{Henry}
Overduin, J.; Coplan, M.; Wilcomb, K.; Henry, R.C.
Curvature Invariants for Charged and Rotating Black Holes.
\emph{Universe} \textbf{2020}, \emph{{6}}, 22, doi:10.3390/universe6020022
\bibitem{14} Henry, R.C. Kretshmann Scalar for a Kerr-Newman Black Hole. \emph{Astrophys. J.} \textbf{2000},~\emph{535},~350--353.
\bibitem{Baker}
Baker, J.G.; Campanelli, M.
Making use of geometrical invariants in black hole collisions.
\emph{Phys.~Rev.~D} \textbf{2000},~\emph{{62}}, 127501, doi:10.1103/PhysRevD.62.127501. 
\bibitem{Abdelqader:2014vaa}
Abdelqader, M.; Lake, K.
Invariant characterization of the Kerr spacetime: Locating the horizon and measuring the mass and spin of rotating black holes using curvature invariants.
\emph{Phys.~Rev.~D} \textbf{2015}, \emph{{91}}, 084017,
doi:10.1103/PhysRevD.91.084017.
\bibitem{13} MacCallum, M.A.H. Spacetime Invariants and Their Uses. \emph{arXiv}~\textbf{2015}, arXiv:1504.06857v1.
\bibitem{16} 
Brooks, D.; MacCallum, M.A.H.; Gregoris, D.; Forget, A.; Coley, A.A.; Chavy-Waddy, P.~C.; McNutt, D.D.
Cartan Invariants and Event Horizon Detection, Extended Version.
\emph{Gen.~Relat.~Gravit.}~\textbf{2018}, \emph{{50}}, 37.
\bibitem{Stephani:2003tm} 
Stephani, H.; Kramer, D.; MacCallum, M.A.H.; Hoenselaers, C.; Herlt, E.
\emph{Exact Solutions of Einstein's Field Equations}; Cambridge University Press: Cambridge, UK, {2003}, doi:10.1017/CBO9780511535185.



\bibitem{ADM} 
Arnowitt, R.L.; Deser, S.; Misner, C.W.
Dynamical Structure and Definition of Energy in General Relativity.
\emph{Phys.~Rev.~} \textbf{1959}, \emph{{116}}, 1322,
doi:10.1103/PhysRev.116.1322
\bibitem{Marqu} 
Marquet, P. The Generalized Warp Drive Concept in the EGR Theory. \emph{Abraham Zelmanov J.} \textbf{2009},~\emph{2},~261--287.
\bibitem{8} Bronnikov, K.A. 1973 Scalar-tensor theory and scalar charge. \emph{Acta Phys. Pol.} \textbf{2009},~\emph{B4},~251--266.
\bibitem{Barcelo:2002} 
Barcelo, C.; Visser, M.
Twilight for the energy conditions?
\emph{Int.~J.~Mod.~Phys.~D} \textbf{2002}, \emph{{11}}, 1553,
doi:10.1142/S0218271802002888
[gr-qc/0205066].
\bibitem{LoboVisser}
Lobo, F.S.N.; Visser, M.
Fundamental limitations on 'warp drive' spacetimes.
\emph{Class. Quantum Gravity} \textbf{2004},~\emph{{21}}, 5871--5892,
doi:10.1088/0264-9381/21/24/011.
\bibitem{Woods}
Woods, R.C.; Baker, R.M.L.; Li, F.; Stephenson, G.V.; Davis, E.W.; Beckwith, A.W.
A new theoretical technique for the measurement of high frequency relic gravitational waves.
\emph{J. Mod. Phys.} \textbf{2011}, \emph{2},~498--518, doi:10.4236/jmp.2011.26060
\end{thebibliography}
\end{document}